# Anatomy of a Crash

Aude Marzuoli, Emmanuel Boidot, Eric Feron
Georgia Institute of Technology
Paul B.C. van Erp, Alexis Ucko, Alexandre Bayen, Mark Hansen
University of California at Berkeley

October 15, 2014

**Abstract**

Transportation networks constitute a critical infrastructure enabling the transfers of passengers and goods, with a significant impact on the economy at different scales. Transportation modes, whether air, road or rail, are coupled and interdependent. The frequent occurrence of perturbations on one or several modes disrupts passengers' entire journeys, directly and through ripple effects. The present paper provides a case report of the Asiana Crash in San Francisco International Airport on July 6th 2013 and its repercussions on the multimodal transportation network. It studies the resulting propagation of disturbances on the transportation infrastructure in the United States. The perturbation takes different forms and varies in scale and time frame : cancellations and delays snowball in the airspace, highway traffic near the airport is impacted by congestion in previously never congested locations, and transit passenger demand exhibit unusual traffic peaks in between airports in the Bay Area. This paper, through a case study, aims at stressing the importance of further data-driven research on interdependent infrastructure networks for increased resilience. The end goal is to form the basis for optimization models behind providing more reliable passenger door-to-door journeys.

## 1 Introduction

In 2012, 2.9 billion passengers boarded an airplane, whether for business or leisure, across the world [1]. Yet, air transport is only a portion of the passenger door-to-door journey, which also relies on other modes of transportation, such as rail, road and water. Transportation modes are usually studied separately as if not interacting, although they are intrinsically coupled through passenger transfers. The failure of one mode disrupts the entire passenger journey. Over the past few years, many disruptions have highlighted the rigid structure of transport infrastructures and the potential for perturbations to snowball across multimodal infrastructures. In particular, the failures and inefficiencies of the air transportation system not only have a significant economic impact but they also stress the importance of putting the passenger at the core of the system [2] [3] [4] [5]. Airlines try to increase aircraft utilization to maximize their revenue, and hence shorten time buffers between scheduled arrivals and departures. Delay propagation to later departure flights is therefore more likely. The capacity of the network to absorb disruptions decreases when demand levels get closer to capacity limits. Thus large-scale delays in the system become more frequent. When a disruption occurs, airline schedule recovery tries to maintain operations and get back to schedule as quickly as possible while minimizing additional costs. The different mechanisms they rely on are aircraft swaps, flight cancellations, crew swaps, reserve crews and passenger rebooking. Usually airlines react by solving the problem in a sequential manner. First, infeasibility of the aircraft schedule is examined, then crewing problems, ground problems and last, the impact on passengers. In 2010, the Islandic volcano eruption resulted in the cancellation of more than 100,000 flights, with stranded passengers and their luggage across Europe, scrambling to reach their destination using other modes. Every year in the US, hurricanes, snow storms or pop-up thunderstorms cause massive cancellations and delays in the entire transportation system. As the number of passenger keeps growing [1], congestion and snowball effects threaten the resilience of the whole multimodal transport infrastructure.

The present paper undertakes a clinical study of the Asiana crash in San Francisco airport on July 6th, 2013 and the resulting large-scale multimodal perturbation that propagated on the airside and the landside.

Our objective is to provide the first case study of infrastructure failure leading to a multimodal disturbance on different time frames and scales across various transportation networks. The higher-level goal is to foster a better understanding of multimodal transportation to increase its resilience and the passenger door-to-door journey. Such a case report can provide the first experimental basis upon which several system engineering methods could be applied to improve the entire passenger journey. These methods encompass network science approaches, classical control and optimization techniques for infrastructure networks, queueing systems for traffic management [6], to name a few.

From a network science perspective, much research has focused on examining the structure of each transportation mode. The world transportation industry is a critical infrastructure with a significant impact on local, national and international economies. Guimera et al. [7] find that the cities with the most connections are not always the most central in the network though. Most cities, or nodes, are peripheral, meaning that the majority of their connections are within their own community. The nodes that connect different communities are usually hubs, but not necessarily global hubs. The structure of the airport network has been extensively studied [8], mostly using airline flight frequency. Sridhar et al. were among the first to examine metrics of the US air transportation network [9]. DeLaurentis [10] described system-wide factors and issues that also matter in the overall system performance, such as the service network topology, economic policy or airline fleet mixture. Conway [11] and Holmes [12] advocate that network science offers a new perspective on air transportation. Because of the network structure, some local delays can have ripple effects on the entire NAS and cause major delays [13]. Flight delays do not accurately reflect the delays imposed upon passengers' full multi-modal itinerary. The growing interest to measure ATM performance calls for metrics, that reflect the passenger's experience. Cook and al. [14] design propagation-centric and passenger-centric performance metrics, and compare them with existing flight-centric metrics.

To the best of our knowledge, there is little work on network coupling or interdependencies and hardly any on transportation infrastructures. One of the most striking examples to date is the electrical blackout in Italy in September 2003: the shutdown of power stations directly led to the failure of nodes in the Internet communication network, which in turn caused further breakdown of power stations [15]. At the theoretical level, the robustness of interdependent random networks is beginning to be understood [16] but research on real-world applications is lacking. Interdependent infrastructure networks are complex cyberphysical systems, and may also be seen as highly optimized tolerant systems [17]. Understanding the observability and controllability of complex networks [18] is critical to ensure their robustness under perturbations.

On the transportation side, there has been extensive research on disturbance propagation in the airspace [19] [20] [21] [22], the impact of airline scheduling of aircraft and crew [23] and the best recovery optimization schemes [24] [25]. Recently, a shift towards passenger-centric metrics in air transportation, as opposed to flight-centric, has been promoted, highlighting the disproportionate impact of airside disruptions on passenger door-to-door journeys [14] [26] [27] [28]. Indeed, disrupted passengers, whose journey was interrupted, only account for 3% of the total passengers, but suffer 39% of the total passenger delay.

The paper is organized as follows. Section 2 provides a brief description of the ASIANA crash and the subsequent events at San Francisco airport. Section 3 evaluates the direct impact of the crash in the Bay Area, both on the air and the ground sides. Section 4 analyzes the propagation of disturbances on the multimodal network, consisting of the air transportation, highway and transit networks. Section 5 concludes the paper and presents future research perspectives.

# 2 Crash description

First let us provide a brief summary of the Asiana crash at San Francisco airport. The layout of San Francisco International Airport (SFO) is displayed in Figure 1. It is the seventh busiest airport in the United States, with about 400,000 movements and 45 millions passengers per year.

On July 6th, 2013, the weather was good, the winds were light. The instrument landing system's vertical guidance (glide slope) on runway 28L was, as scheduled, out of service. At 11:28 a.m, Asiana Airlines Flight 214, a Boeing 777-200 ER aircraft, crashed just short of runway 28L's threshold at San Francisco International Airport. Of the 307 people aboard, 3 died, 181 others were injured. The crash resulted in a five hour total closure (and cancellation/redirection of all fights) of the runways at the airport. By 3:30 p.m. PDT, the two runways perpendicular to 28L were reopened; runway 10L/28R (parallel to the runway of the accident) remained closed for more than 24 hours. The accident runway, 10R/28L, reopened on July 12.

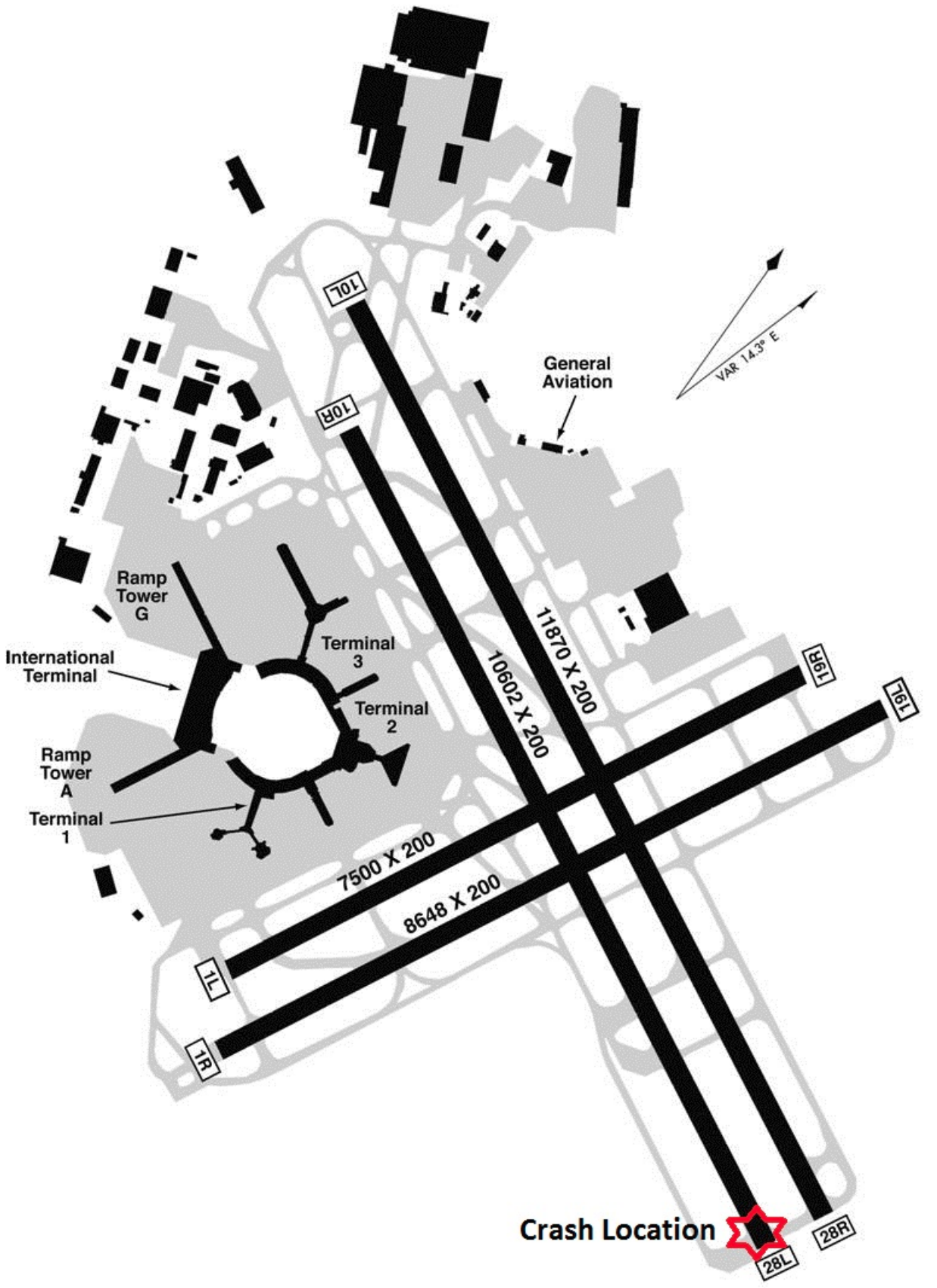

Figure 1: San Francisco Airport Layout.

The National Transportation Safety Board provides the accident timeline, as illustrated in Figure 2.The accident investigation submission [29] states that "the probable cause of this accident was the flight crew's failure to monitor and maintain a minimum airspeed during a final approach, resulting in a deviation below the intended glide path and an impact with terrain. Contributing to this failure were (1) inconsistencies in the aircraft's automation logic, which led the crew to believe that the autothrottle was maintaining the airspeed set by the crew; and (2) autothrottle logic that unexpectedly disabled the aircraft's minimum airspeed protection." The significant contributing factors to the accident identified are : "(1) inadequate warning systems to alert the flight crew that the autothrottle had (i) stopped maintaining the set airspeed and (ii) stopped providing stall protection support; (2) a low speed alerting system that did not provide adequate time for recovery in an approach-to-landing configuration; (3) the flight crew's failure to execute a timely go-around when the conditions required it by the company's procedures and, instead, to continue an unstabilized approach; and (4) air traffic control instructions and procedures that led to an excessive pilot workload during a high-energy final approach."

# 3 Airside Analysis

The crash led to the closure of SFO and, even after the airport reopened, its capacity was reduced significantly. The crash led to cancellations, diversions and delays at SFO, and impacted the rest of the airspace with ripple effects. The work presented is based upon publicly available data from the Bureau of Transportation Statistics that are primarily used to evaluate airline on-time performance.

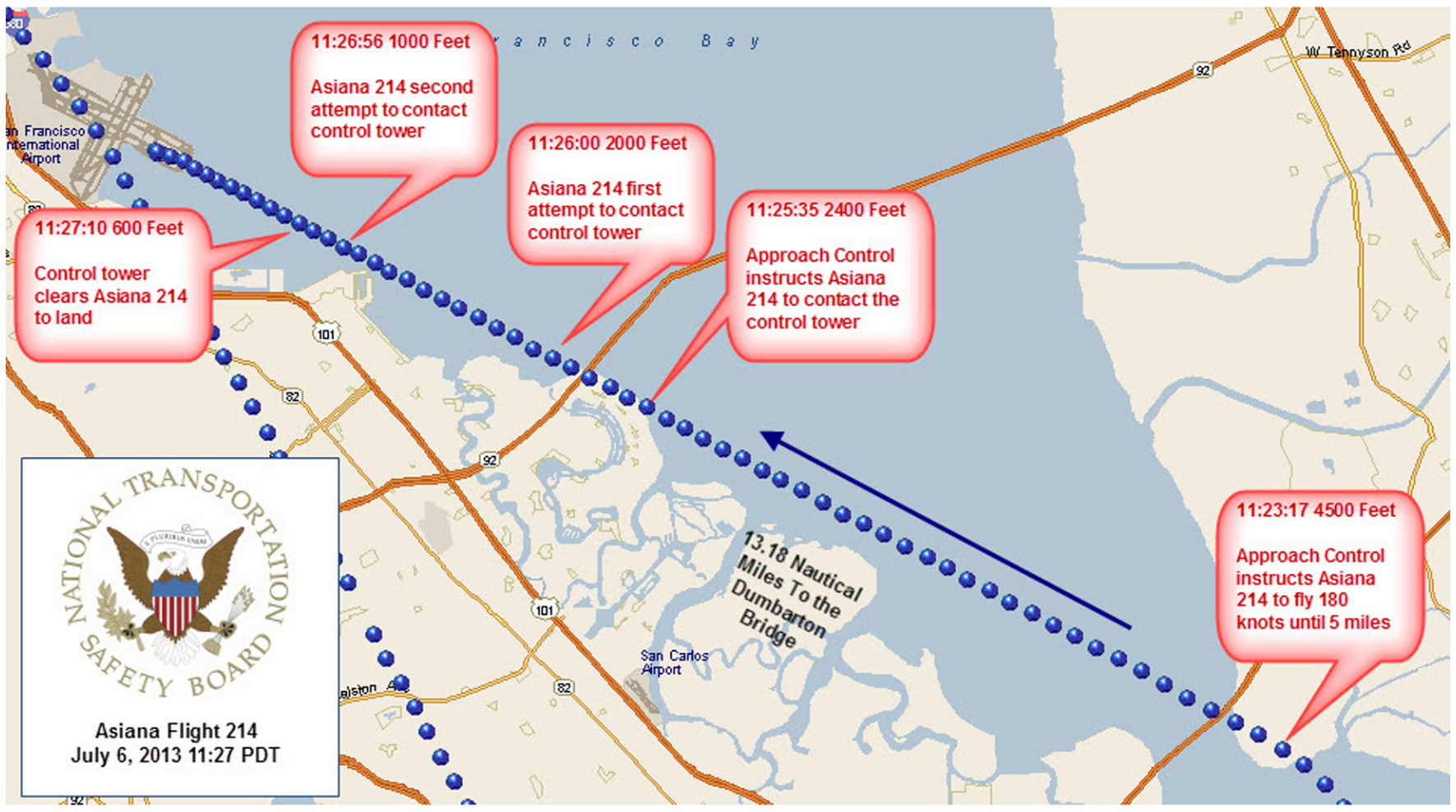

Figure 2: Asiana 214 Approach phase. The blue bullets represent the aircraft trajectories. The information in red describes the exchanges between the Asiana 214 flight and Air Traffic Control.

## 3.1 Impact of the crash in San Francisco

### 3.1.1 Departures, Arrivals, Cancellations and Diversions at SFO

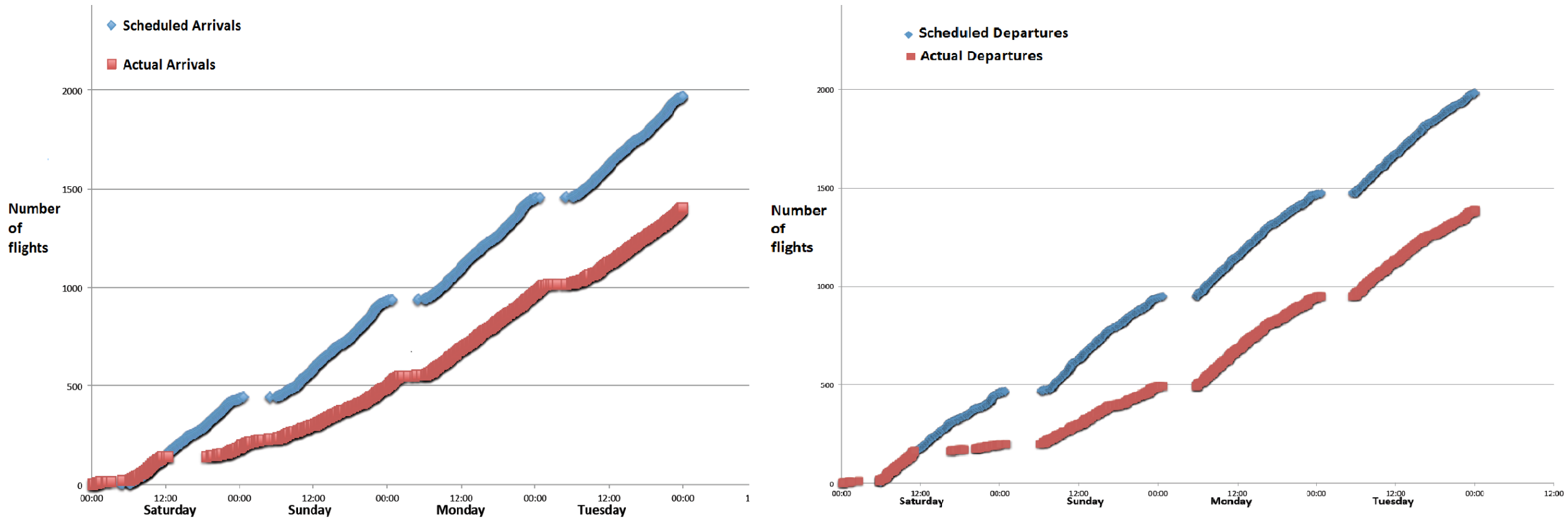

Figure 3: Scheduled vs Actual Arrivals and Departures at SFO airport, July 6th-9th 2013.

Figure 3 represents the difference between scheduled and actual operations at SFO from Saturday, July 6th 2013 to Tuesday, July 8th 2013. The divergence between scheduled and actual departures, as well as scheduled and actual departures begins immediately after the crash. The airport is closed until the two shorter runways, perpendicular to the crash runway, reopen in the afternoon. Departures and arrivals then resume at a slower pace than usual because of reduced runway capacity at the airport. Summing the results over four days, more than 660 flights scheduled to land at SFO airport had either been canceled or diverted, and more than 580 flights had been canceled or diverted at departure from SFO.

Figure 4 displays the temporal evolution of diversions and cancellations to or from SFO airport from Saturday July 6th 2013 to Tuesday July 9th. First Figure 4 shows that diversions mostly occurred on Saturday as well as on Sunday. There are several departure diversions, meaning that flights that departed from SFO made a stop before reaching their final destination, mostly on Saturday evening and Sunday morning, when there are fewer arrival disruptions. The proportion of diversions is high : 17% of arrival flights to SFO were diverted on Saturday. After Sunday the number of diversions went back to normal, while cancellations remained considerable. Indeed, cancellations span the four days without any noticeable pattern regarding their timing. A closer look at the repartition of cancellations and diversions over the crash week-end in Tables 5 highlights the impact of the crash. More than half of the scheduled departures and almost half the scheduled arrivals were canceled on Saturday, and the figures slowly decreased until Tuesday.

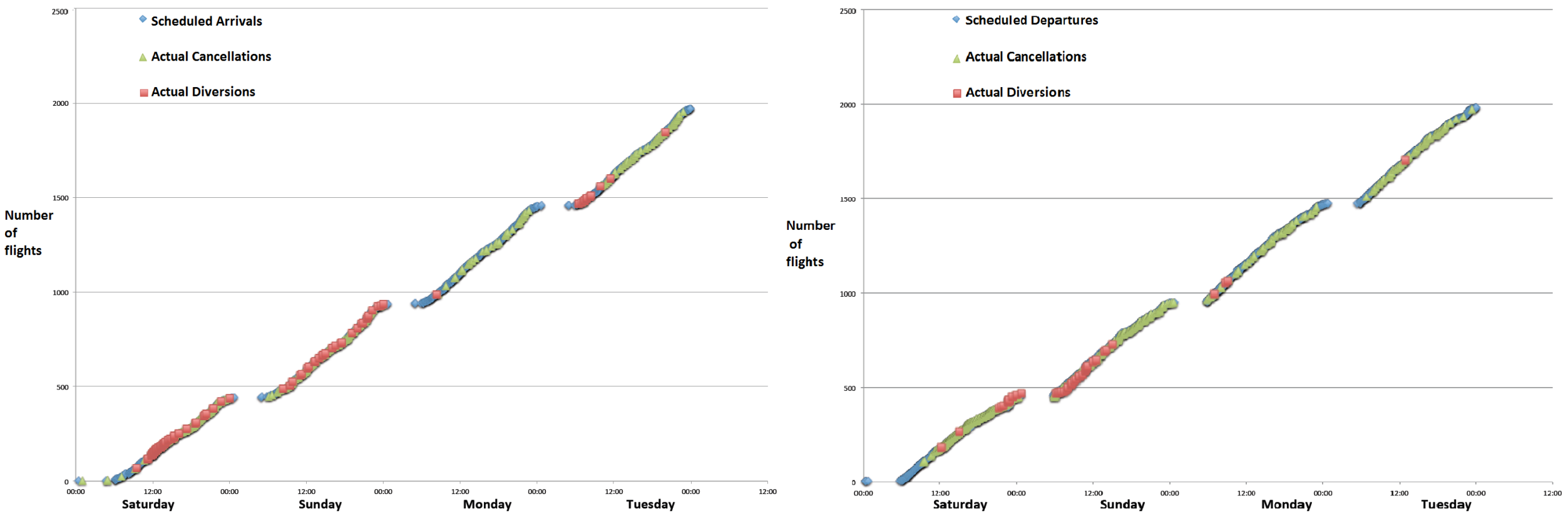

Figure 4: Temporal evolution of Cancellations and Diversions at SFO airport, July 6th-9th.

| Day | Complete | Cancelled | Diverted |
|---|---|---|---|
| Sat | 185 | 180 | 74 |
| Sun | 310 | 155 | 30 |
| Mon | 476 | 43 | 1 |
| Tue | 433 | 70 | 14 |

(a) Arrivals in SFO

| Day | Complete | Cancelled | Diverted |
|---|---|---|---|
| Sat | 198 | 231 | 11 |
| Sun | 293 | 174 | 30 |
| Mon | 456 | 62 | 2 |
| Tue | 439 | 74 | 1 |

(b) Departures from SFO

Figure 5: Number of complete, canceled and diverted flights to and from SFO during the entire crash week-end

| Day | DEN | LAS | LAX | MSP | OAK | PHX | RNO | SJC | SLC | SMF | Sum |
|---|---|---|---|---|---|---|---|---|---|---|---|
| Sat | 4 | 7 | 8 | 1 | 15 | 3 | 3 | 22 | 1 | 10 | 74 |
| Sun | 0 | 0 | 0 | 0 | 19 | 0 | 0 | 11 | 0 | 0 | 30 |
| Mon | 0 | 0 | 0 | 0 | 1 | 0 | 0 | 0 | 0 | 0 | 1 |
| Tue | 0 | 0 | 0 | 0 | 4 | 0 | 0 | 7 | 1 | 2 | 14 |

Figure 6: Number of flights which were supposed to land at SFO airport and were diverted to other airports

Operations were worse on Tuesday, July 9th than on Monday, July 8th. Moreover, due to the closure of the crash runway, runway capacity was still significantly reduced, leading to many cancellations. There are very few diversions after Sunday. This is to be expected since diversions are usually tactical operations. Upon further investigation of the departure diversions on Saturday evening and Sunday morning, these diversions impacted medium-haul flights only, with a short stop in SLC airport and reached their final destination with little delay. The most likely explanation is that these flights were performed by fairly heavy aircraft. Because only the two shorter runways were opened until Sunday afternoon, they probably had to depart with less fuel than needed for their entire trip and their planned refueling at another airport appears in the data as a diversion.

The major carriers flights were diverted to a number of airports, as reported in the BTS data. The other Bay Area airports, Oakland (OAK) and San Jose (SJC) accommodated most flights, from Saturday to Tuesday. Nevertheless, several other airports, as far as Denver, Los Angeles and Las Vegas, welcomed many diverted flights on the crash day. Figure 7 displays the estimated number of passengers who were diverted to different airports than SFO from July 6th to July 9th, based on the load factor reported by each airline for July 2013 to the BTS. Figure 7 shows that several thousands of passengers were diverted to other airports from July 6th to July 9th. The BTS data does not provide indications regarding diversions of international flights but news reports [30]that several international flights were diverted to Seattle Tacomac (SEA) on

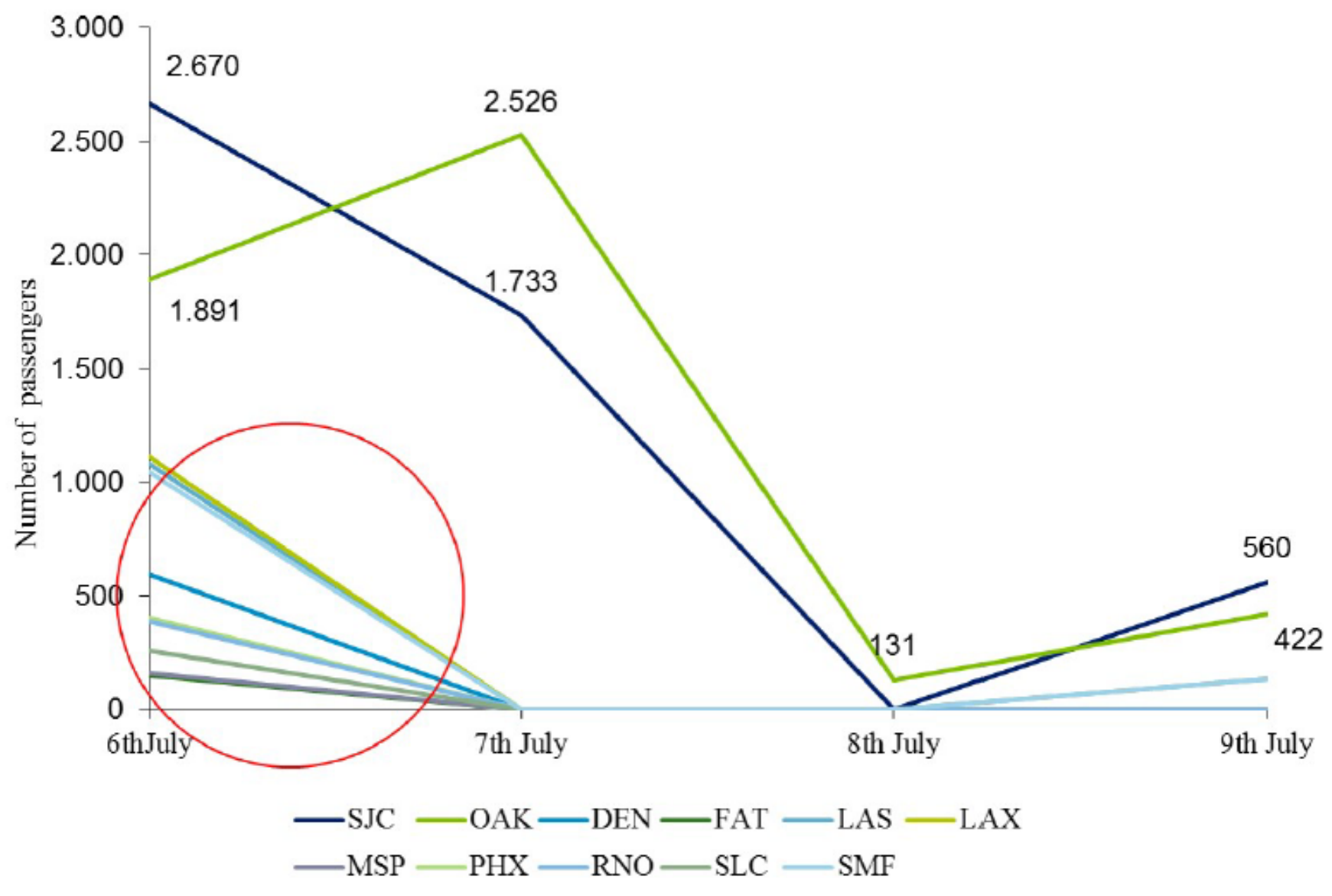

Figure 7: Estimated number of passengers diverted to different airports from July 6th to July 9th 2013. The red circle corresponds to the crash day during which there were the most diversions.

Saturday, July 6th, coming from London, Dubai, Frankfurt, Paris and Zurich.

Many more issues arose when flights were diverted to airports in which their carrier does not operate. For instance, a SFO-bound United Airlines flight from Seattle was diverted to Oakland. Local news reporters [31] interviewed the 6th of July 2013 some of the flight passengers: "United has no support here. They sent a dislocation team, but basically what they keep saying is: "You're dislocated." " The officials said they had to bring extra staff to accommodate passengers who were landing at the same time. Moreover, many passengers were diverted to airports where their airline operates at low frequency.

### 3.1.2 Delays at SFO airport

When it comes to operations at the airport itself, Figure 8 displays the delay minutes for each departing and arrival flight against their scheduled departure or arrival time at SFO airport during (i) the crash day and (ii) Saturday July 27th, which is used as a reference day. There were 879 scheduled flights at SFO on July 6th and 901 on July 27th, corresponding to domestic US carriers. On July 6th, there were a total of 411 cancellations and 85 diversions, whereas on July 27th, 8 flights were cancelled and none diverted. Immediately after the crash, departure and arrival delays rise significantly and are much higher than on July 27th, although the number of operations is considerably smaller. The delays go back to almost normal levels after 10 pm. Figure 9 presents the taxi-out and taxi-in times at SFO airport through the crash day and a normal Saturday. Contrary to departure and arrival delays, the taxi-out times were completely normal through the day. This means that the departure delay observed is primarily due to delay incurred at the gate. Taxi-in times were normal except around the crash time. Because of the number of emergency vehicles going to the crash runway, arrival flights on the ground may have been held to let them through.

## 3.2 Impact of the crash on the Air Transportation Network

Cancellations and delays due to the crash at SFO rippled through the airspace and lasted several days.

### 3.2.1 Cancellations and their propagation

Figure 10 shows the number of departure and arrival cancellations at SFO for the entire month of July. The day of the crash, Saturday, is the worst in terms of cancellations, with more than 45% of the scheduled flights cancelled. Sunday July 7th is the second worst. The recovery takes more than a week after the crash, with the week from July 8th to July 12th witnessing cancellations of more than 10% of the number of scheduled flights each day.

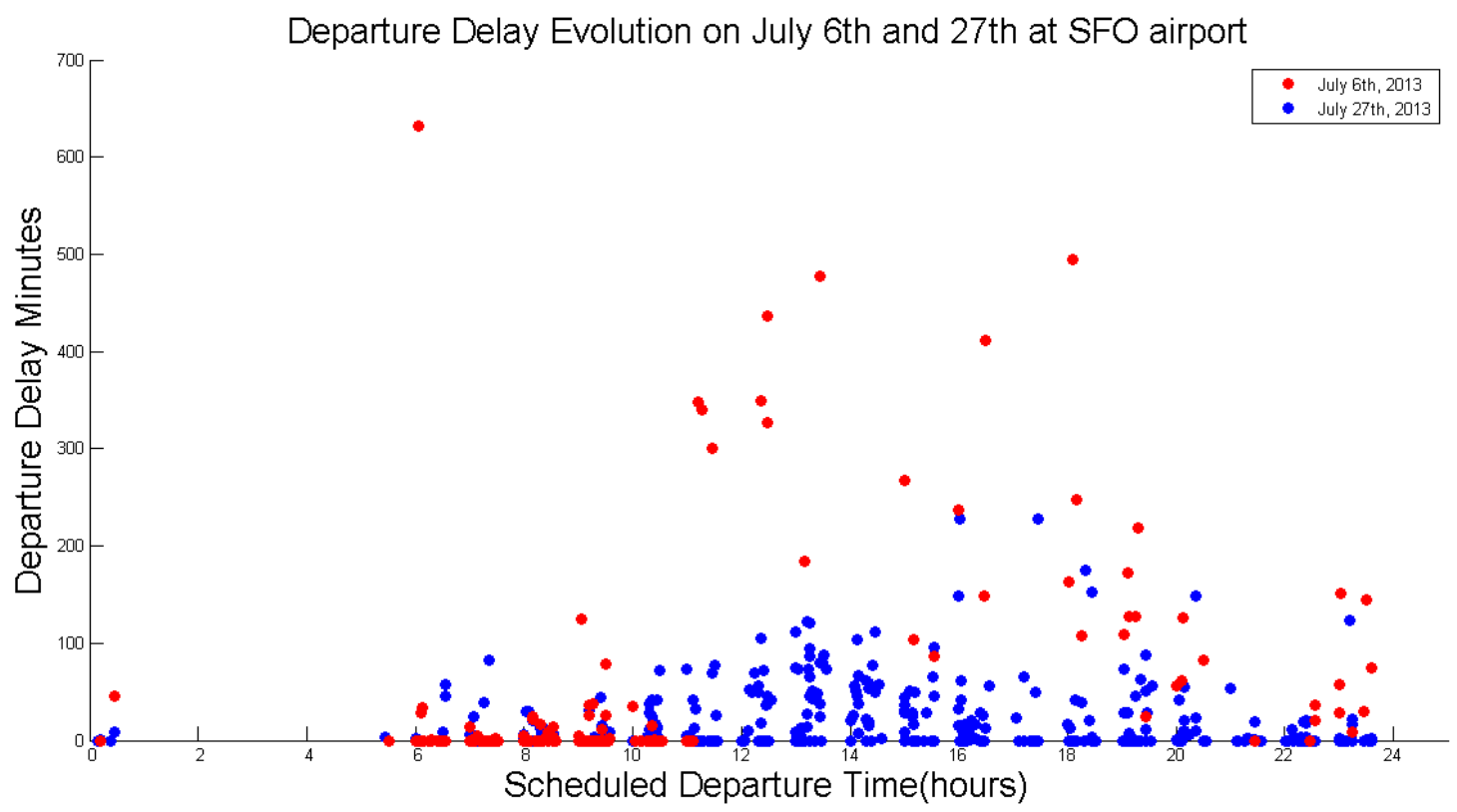

(a) Departure Delay

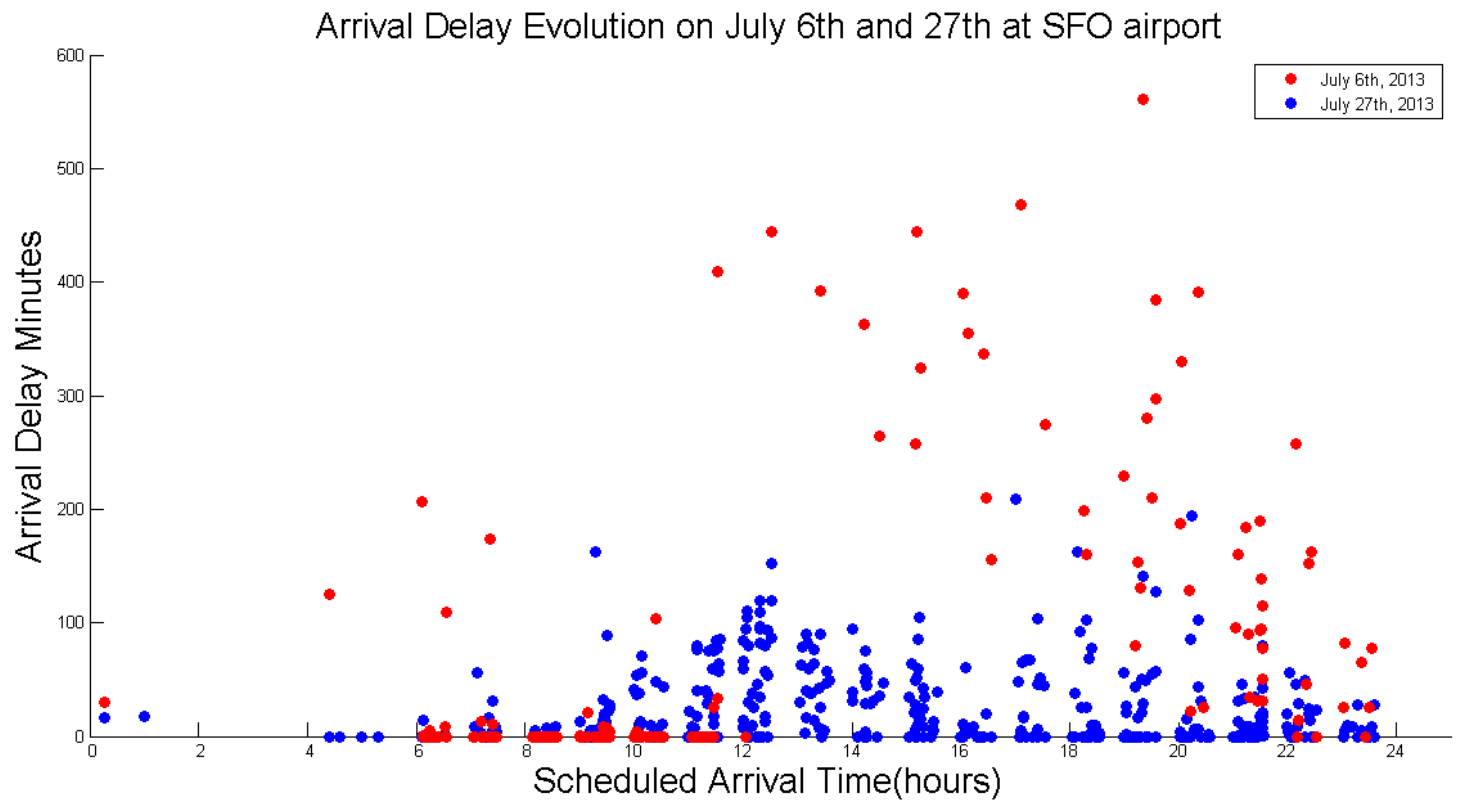

(b) Arrival Delay

Figure 8: Delay comparison between the crash day and a normal Saturday at SFO.

Figure 11 shows the number of departure and arrival cancellations are Los Angeles International Airport (LAX) and Seattle-Tacoma International Airport (SEA). Among the other top 30 airports in the US, LAX and SEA were most affected by cancellations due to the crash. On July 6th and 7th, the proportion of cancelled flights at LAX and SEA was highest for the month of July, with more than 5% of cancelled flights at LAX and 3% at SEA.

Cancellations can propagate through schedules. Indeed, a given aircraft is scheduled to fly several legs through a given day. Once one of these flights has been canceled, the airline tries to get back on schedule, but this schedule recovery is airline- and aircraft-specific. To analyze this propagation phenomenon, the tail numbers of all aircraft involved with flights canceled at departure or arrival to SFO airport from July 6th to July 9th were tracked. For the available tail numbers, each aircraft's individual schedule is recovered. The number of flights each aircraft was supposed to fly is computed. Among these scheduled flights, the total number of cancellations is recorded. In Table 2, the number of cancellations that aircraft without and with tail numbers encounter are summed. It provides the total number of cancellations directly attributable to the SFO crash over the crash week-end. The total number of cancellations regarding flights departing or arriving at one of the top 35 airports in the US is also computed for the crash week-end. In the BTS data, some tail numbers are missing, making these aircraft impossible to track. Such flights are counted in Table 1 under 'missing aircraft id'. The ratio between cancellations attributable to the crash and cancellations in the entire airspace is underestimated, because of missing tail numbers. On the day of the crash, the propagation of cancellations due to the Asiana crash accounts for more than 85% of all cancellations in the airspace, more than 50% on Sunday and more than 25% on Monday and Tuesday. Over the four days, the Asiana crash led

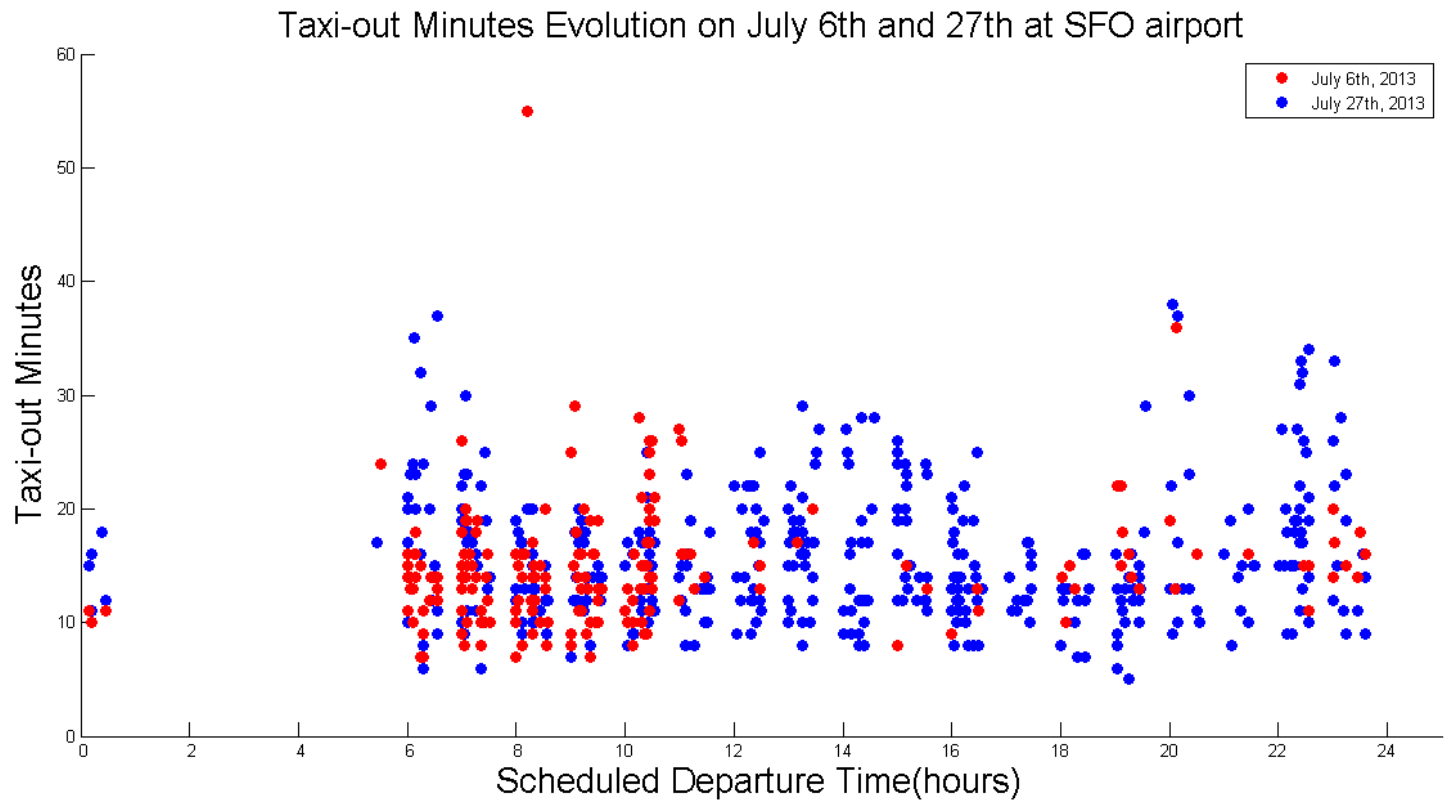

(a) Taxi-out time

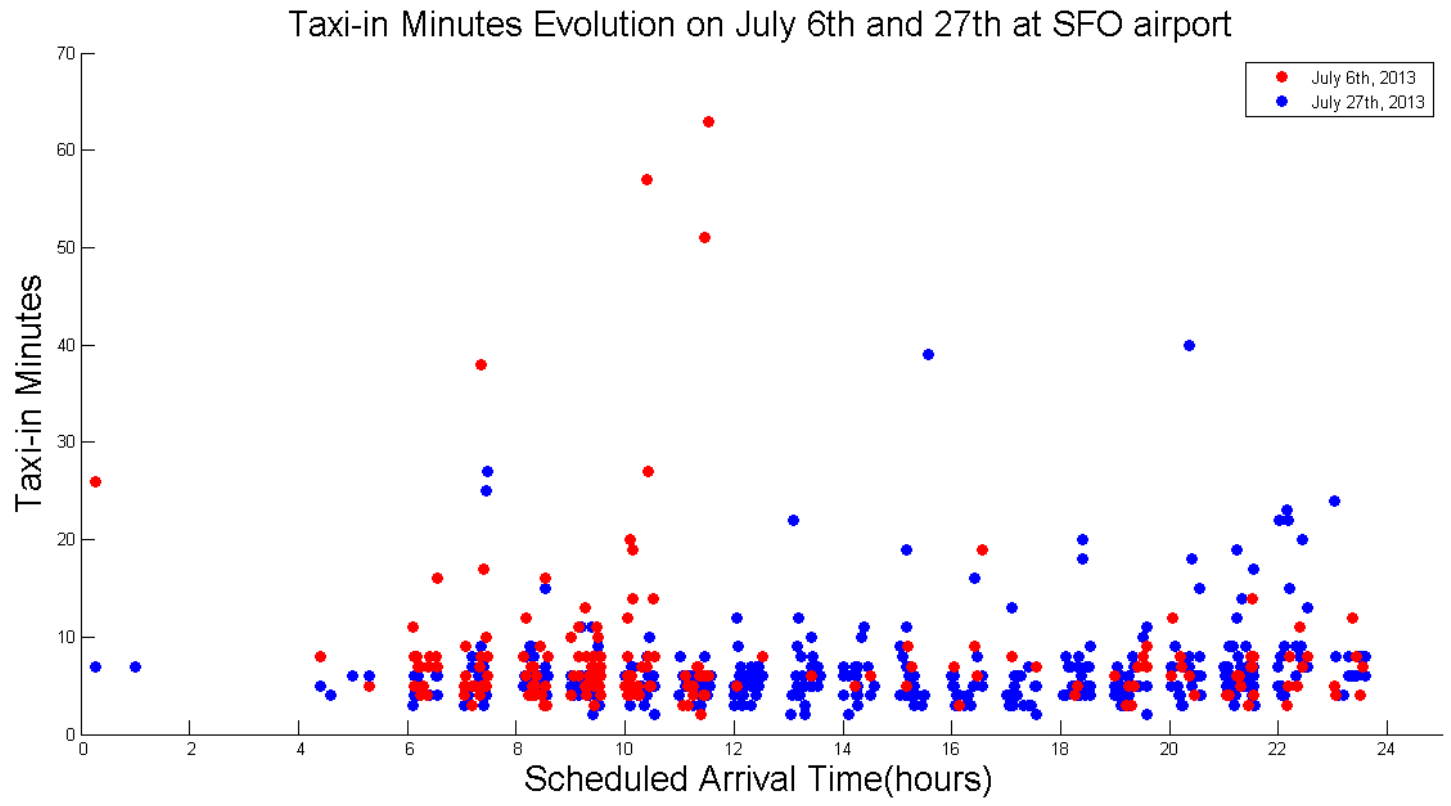

(b) Taxi-in time

Figure 9: Taxi time comparison between the crash day and a normal Saturday at SFO.

to more than 49% of all cancellations in the US airspace.

Table 1: Flight cancellations propagation due to SFO airport.

| Day | Number of missing aircraft id | Number of available aircraft id | For the present aircraft id, number of flights scheduled | For the present aircraft id, number of flights cancelled |
|---|---|---|---|---|
| July 6th | 144 | 137 | 707 | 279 |
| July 7th | 62 | 119 | 708 | 269 |
| July 8th | 17 | 48 | 338 | 92 |
| July 9th | 11 | 59 | 452 | 139 |

Table 2: Cancellations in the airspace attributable to perturbations at SFO airport.

| Day | Number of flights canceled departing or arriving at one of the top 35 airports in the US | Number of cancellations due to SFO perturbations | Percentage due to SFO |
|---|---|---|---|
| July 6th | 488 | 423 | 86% |
| July 7th | 609 | 331 | 54% |
| July 8th | 456 | 109 | 24% |
| July 9th | 510 | 150 | 30% |

Figure 12 shows, for each aircraft with a tail number that encountered a cancellation to or from SFO

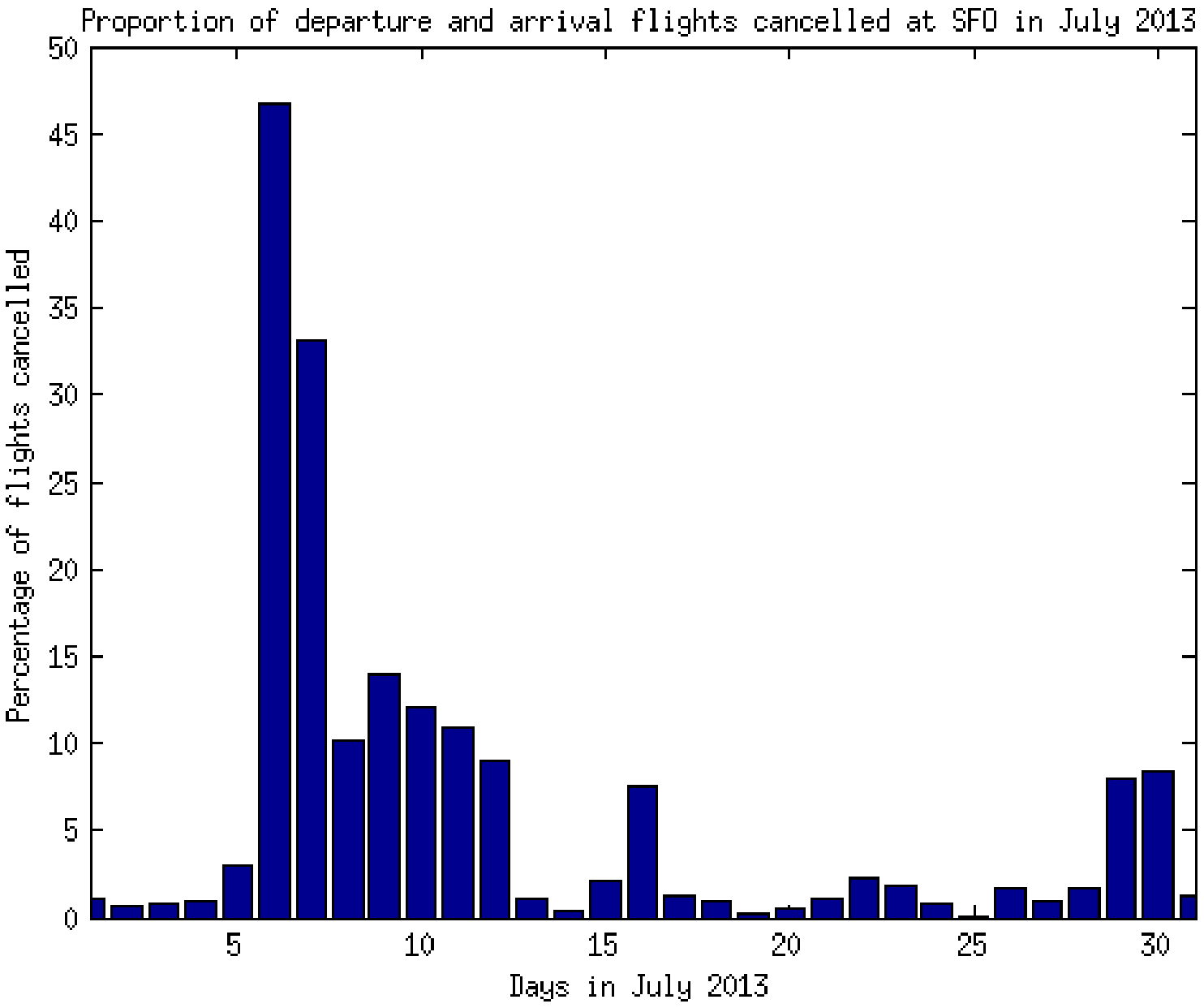

Figure 10: Proportion of cancelled flights among the total scheduled traffic departing or arriving at SFO airport for July 2013.

airport, how much of its schedule was disrupted. Some aircraft were supposed to operate up to eight legs on the crash day. Some aircraft had a first flight cancelled early in their schedule and could not perform any of the remaining legs through the day, whereas others encountered cancellations but could still complete most of their scheduled legs. On Saturday July 6th and Sunday July 7th, most aircraft completed between one and two thirds of their scheduled legs. On Monday July 8th and Tuesday July 9th, more than 50% of the aircraft that encountered a cancellation were then able to complete more than two thirds of their scheduled legs.

#### 3.2.2 Delays and their propagation

To evaluate the impact of the crash on delays throughout the national airspace, the number of delayed aircraft at the top 35 passenger airports in the US is computed for July 2013. The results are displayed in Figure 13. Saturday, July 6th and Sunday, July 7th have some of the lowest total delay in the entire month because cancelled flights are not accounted for in the delays. Since a large proportion of flights were cancelled, even if many of the maintained flights were delayed, the effect of lower flight volume made the overall delay lower.

A visualization tool, inspired from the publicly available "misery map" from Flight Aware [32] was developed to display the proportion of delayed and on-time flights at the top passenger airports in the US over 4-hour periods. The tool also ranks these airports by number of cancellations. Figures 14 and 15 are screen shots of the visualization tool through July 6th and 7th. The time indicated is Pacific time. First, on July 6th, before the crash, Chicago O'Hare was the airport with the most cancellations and the highest proportion of delayed flights, because of a weather perturbation. Right after the crash, the number of cancellations at SFO increases significantly, leading to cancellations at LAX, PHX, SEA in particular. ATL cancellations increase too, but it is also due to the weather pattern observed that day. The proportion of delayed flights also increases throughout the entire airspace. On July 7th, the proportion of delayed flights is much higher than on the previous day at most of the busiest airports, particularly in the afternoon. This could also be an effect of the end of a holiday week-end. For instance, the number of cancellations is much higher in the New York Area airports and Boston than on the previous day.

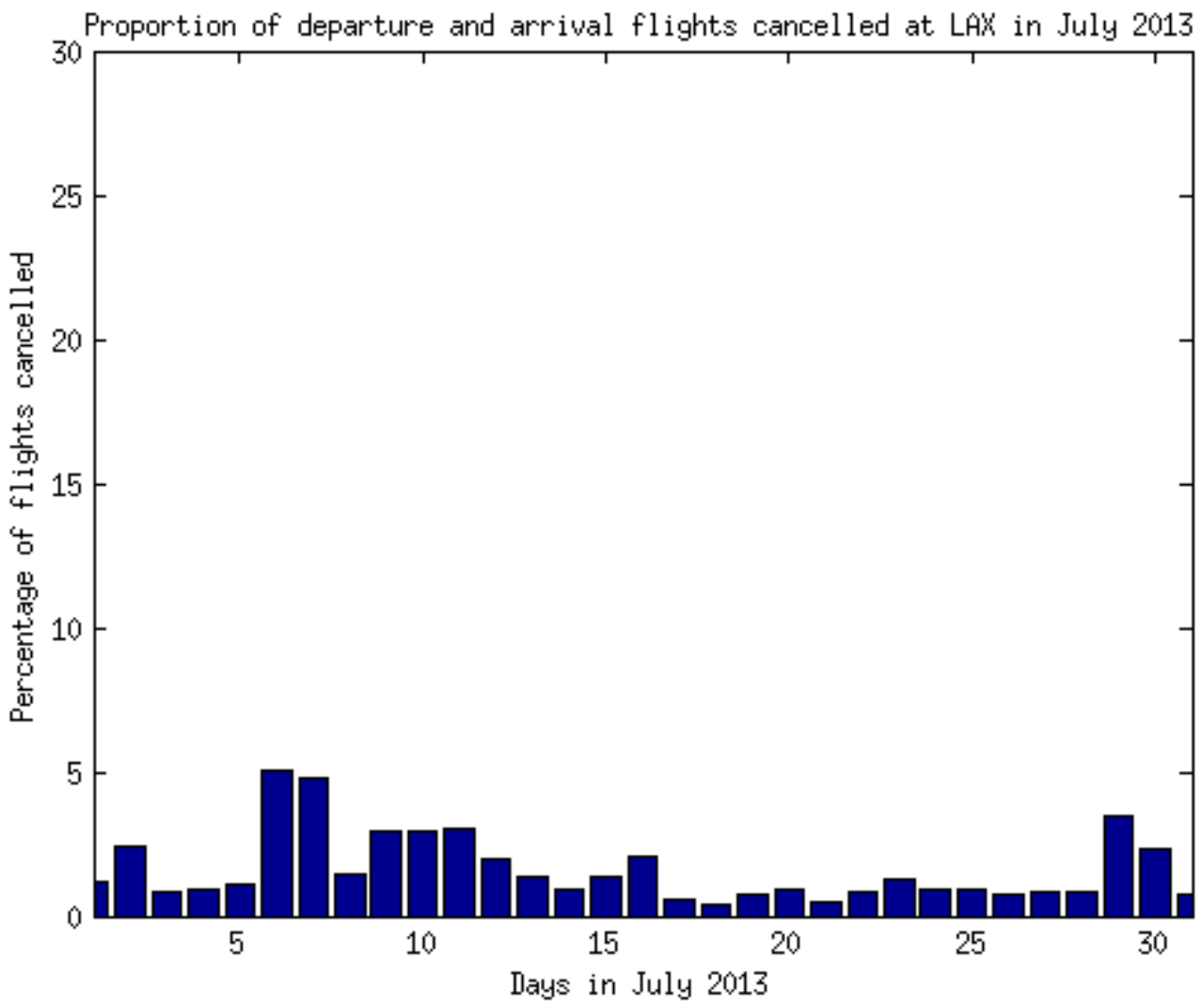

(a) LAX airport

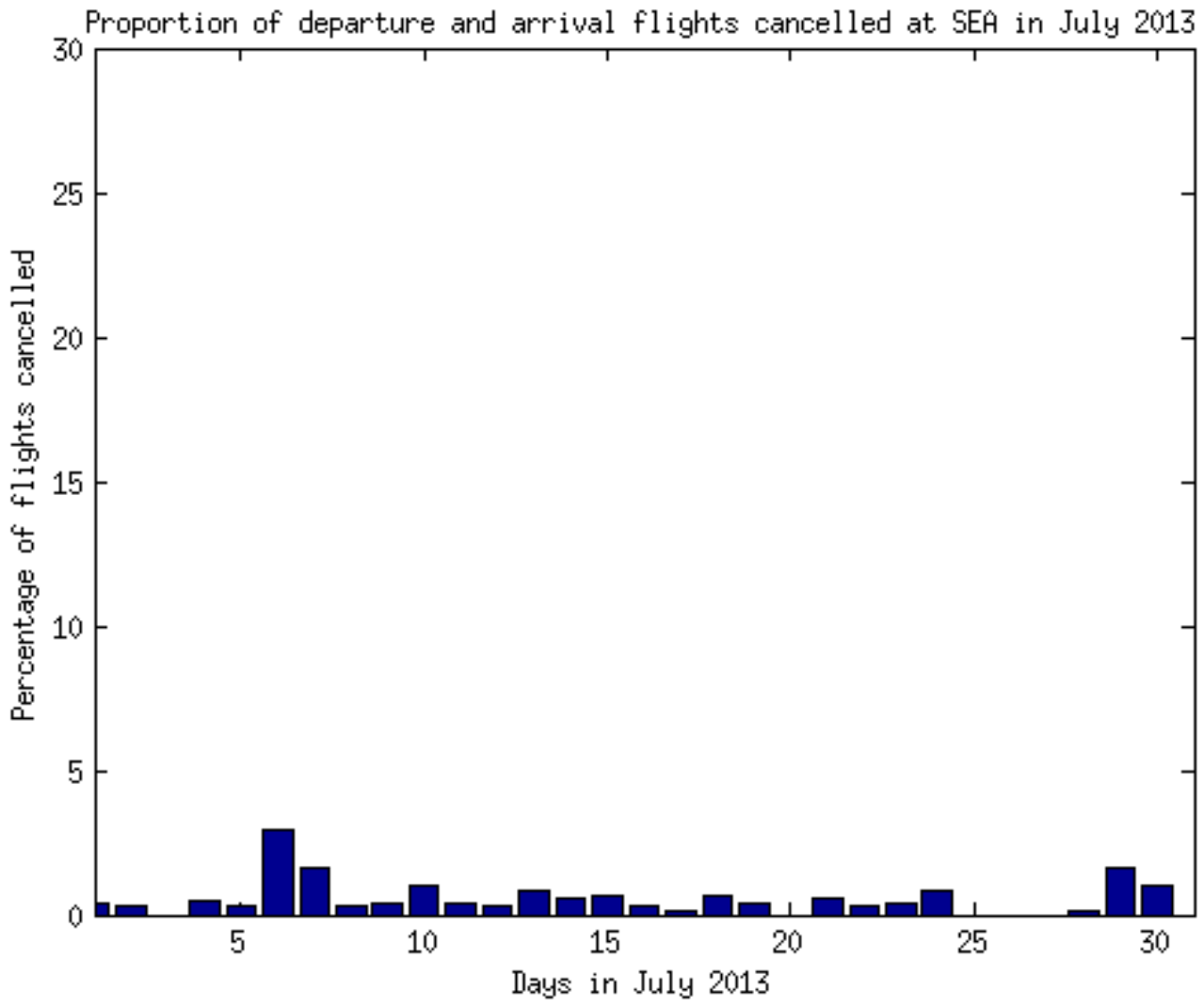

(b) SEA airport

Figure 11: Proportion of cancelled flights among the total scheduled traffic departing or arriving at Los Angeles (LAX) and Seattle (SEA) airports for July 2013.

## 3.3 Cost analysis

The overall cost of the crash is evaluated for the period ranging from Saturday, July 6 to Tuesday, July 9. This cost can be broken down into delay cost, cancellation cost and diversion cost.

### 3.3.1 Delay Cost

Time is a valuable economic resource that may be devoted to work or leisure activities. Because travel takes time, it imposes an opportunity cost equal to the individual value of time in work or leisure activity. Moreover, since travel may take place under undesirable circumstances, including long waits or rides aboard a crowded or uncomfortable vehicle, it may impose an additional cost on travelers. Travel time saved or lost as a result of investments or regulatory actions should be valued in benefit-cost analyses to reflect both the opportunity cost and discomfort, if any, people experience when traveling.

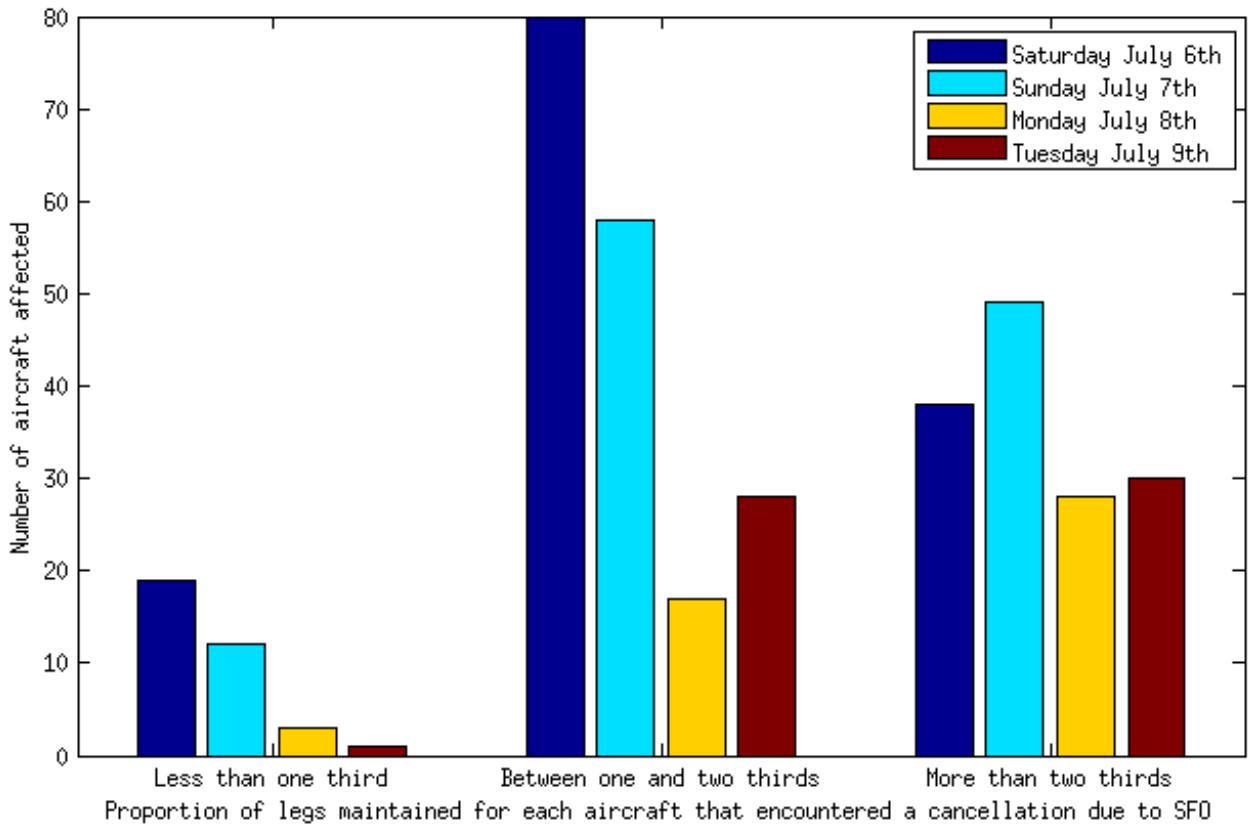

Figure 12: Number of flights maintained per aircraft encountering a cancellation due to SFO airport from Saturday July 6th to Tuesday July 9th.

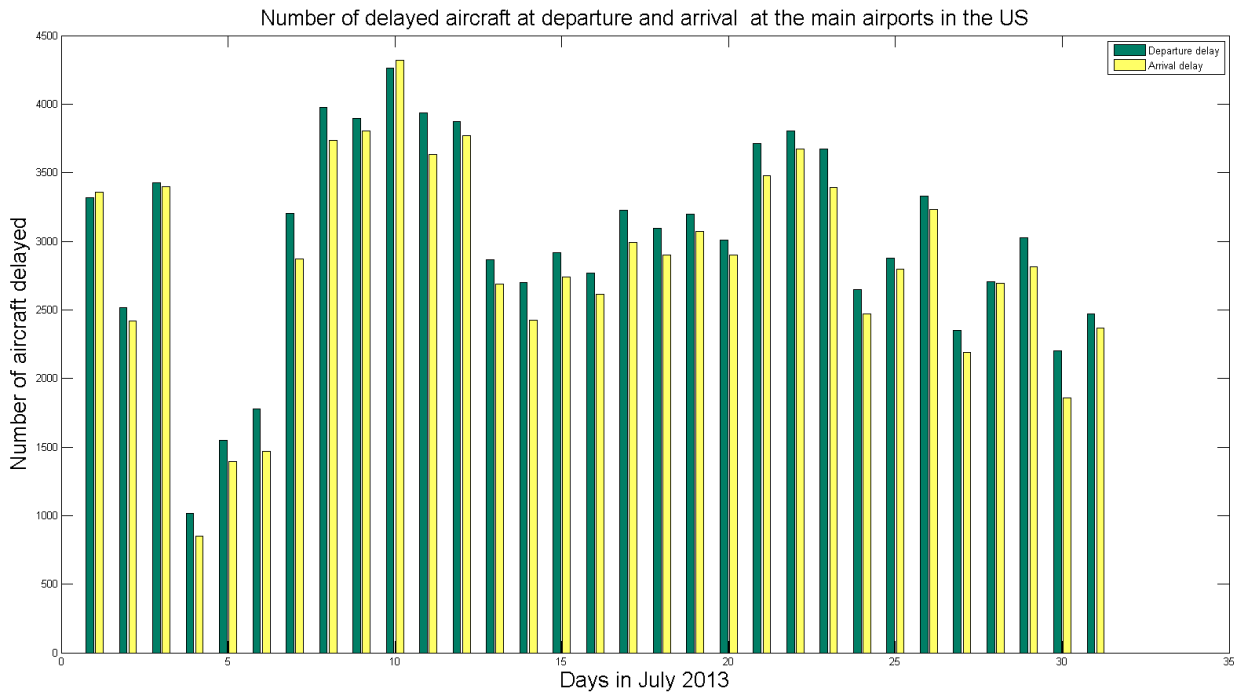

Figure 13: Delays at the top 35 airports for July 2013.

The Department of Transportation (DOT) provides recommended values for aviation passenger travel time [33] on all-purpose air carriers, of \$ 40.10 per hour and person.

First we consider the delayed arrival flights to SFO. The capacity of all the delayed flights during the considered time period is retrieved. The BTS provides the average load factor, 90.51%, for July 2013 for departure and arrival flights at SFO airport. This leads to computing an estimate of the total number of passengers delayed, the number of hours of delay and the associated costs.

$$Total\ Passengers\ per\ Flight = \ Load\ Factor\ \times\ Flight\ Capacity$$

$$Total\ Delays = \sum_{\substack{i=0 \\ i}}^{n} Aircraft\ Delays_i\ \times\ Passengers\ per\ Aircraft_i$$

$$Cost\ per\ Day = \ Total\ Delay\ \times\ Recommended\ Hourly\ Values\ of\ travel\ time\ savings$$

Table 3: Costs of aircraft-borne delays from July 6th to July 9th.

| Days | Cost of delays (\$) |
|---|---|
| July 6 | 837,216 |
| July 7 | 1,608,768 |
| July 8 | 1,772,928 |
| July 9 | 1,890,576 |
| Total | 6,109,488 |

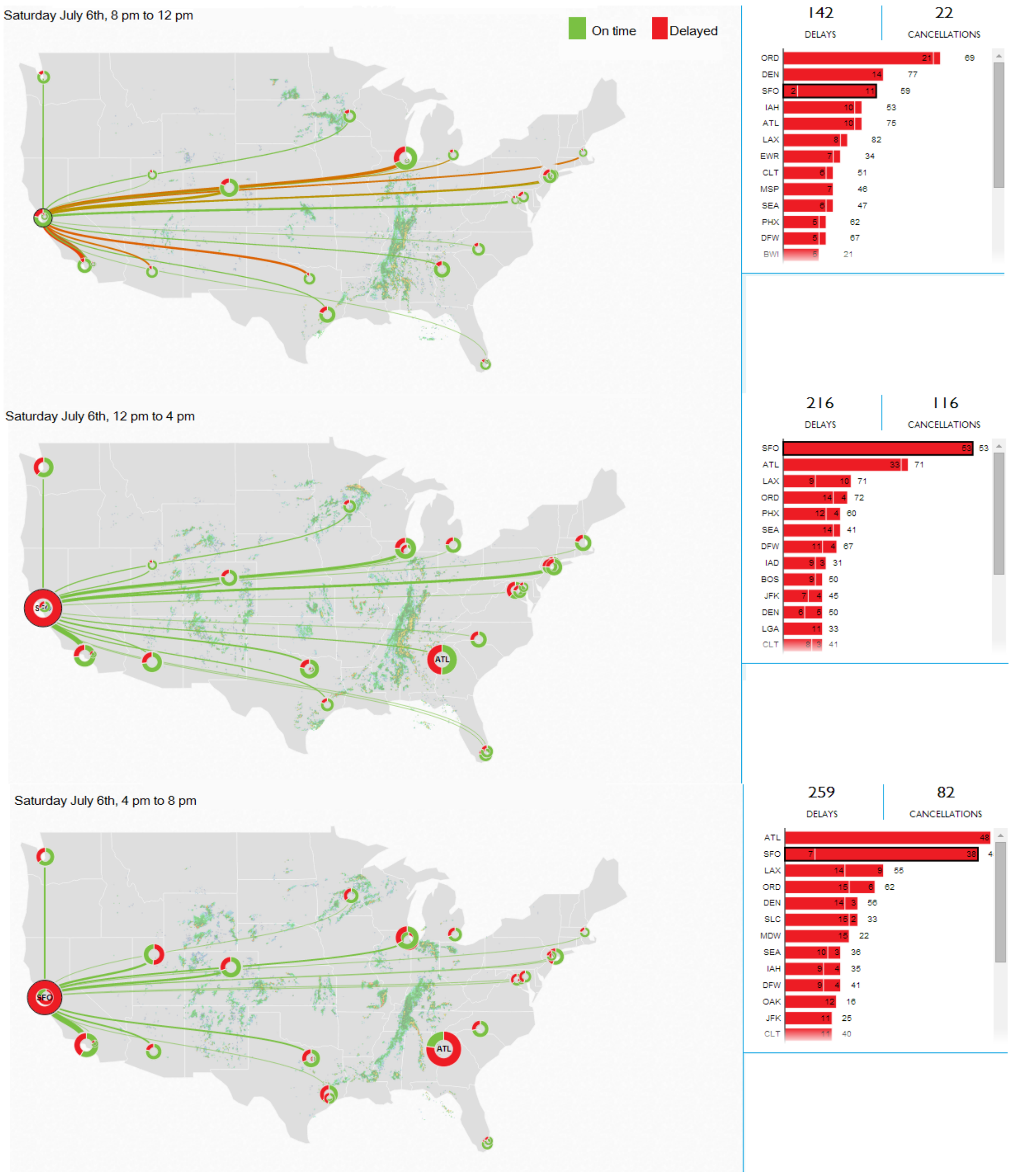

Figure 14: Proportion of delayed flights and number of cancellations at the top airports on July 6 2013.

Because there were more cancelled flights and fewer delayed flights on the crash day compared with Sunday, the delay cost is lower on Saturday. The overall delay cost for arrivals reaches 8 millions \$. The same method is applied to estimate the delay cost for departures from SFO airport. The corresponding load factor is equal to 84.66%. A similar trend on departure flights is observed.

Table 4: Costs of delays from July 6th to July 9th.

| | Cost of delays (million \$) |
|---|---|
| Departures from SFO | 9.5 |
| Arrivals to SFO | 14.1 |
| Total Cost | 23.6 |

Table 4 summarizes the costs of delays during the crash week-end.

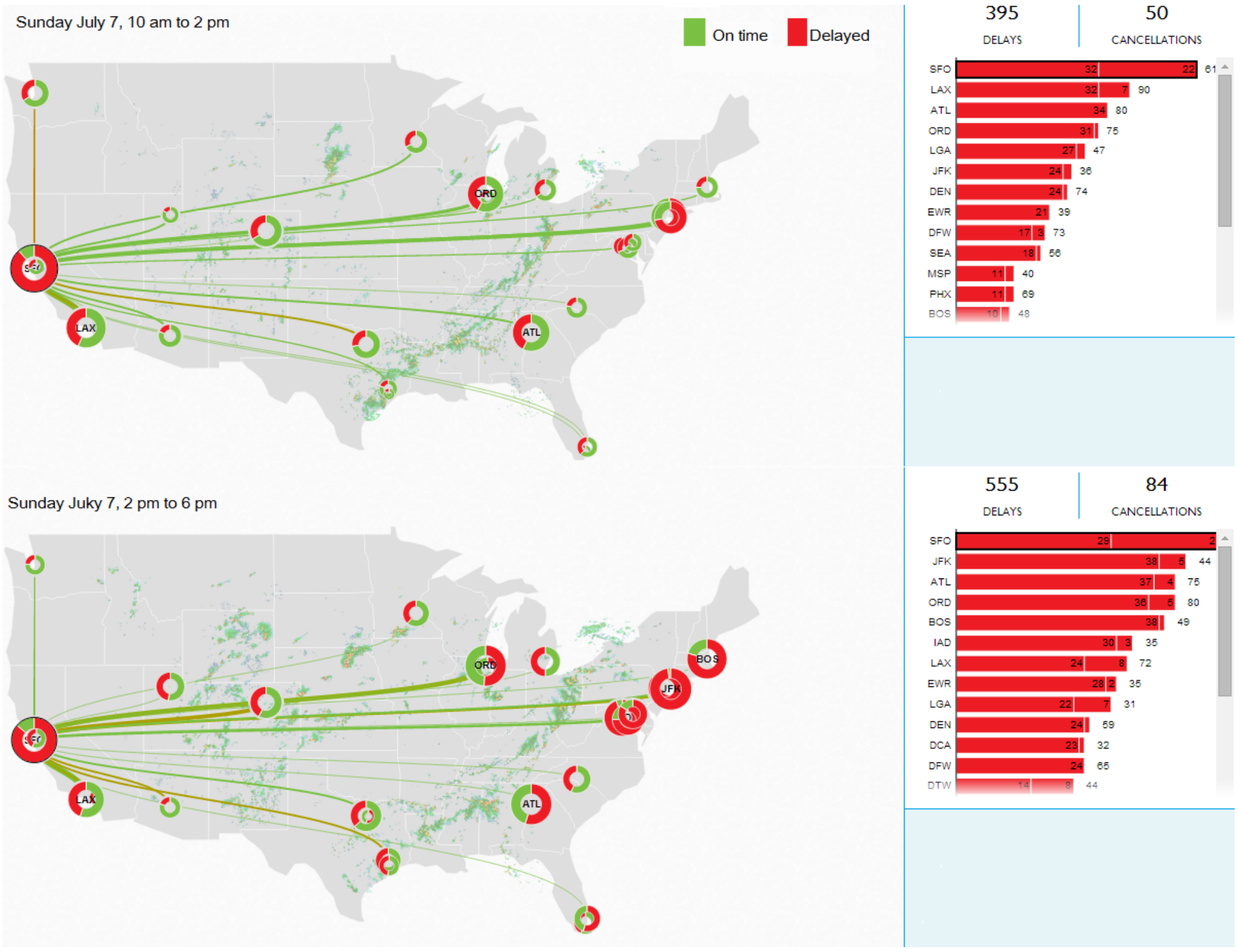

Figure 15: Proportion of delayed flights and number of cancellations at the top airports on July 7 2013.

### 3.3.2 Cancellations

There have been several attempts to estimate the equivalence between cancellations and delay minutes. One solution, used in both academia and industry, is to assign a rough estimate to cancellation cost based on their knowledge and experience. Sridhar [34] states that one cancellation is equivalent to approximately 200-300 minutes of delay. In an unpublished study by Metron Aviation [35], the cost of a flight cancellation was estimated at 6,000 $. In this subsection, we draw a distinction between the airplane standpoint and the passengers standpoint.

**Airplane stand point** The working assumption is that one cancellation is equivalent to 250 minutes of delay. Given the number of cancellations per day, the equivalent delays and the total delays are computed by multiplying the number of total passengers and the delays. The same method is applied to departure flights, and the same trend observed. As for delayed flights, the number of passengers affected is a function of the load factor and the airplane capacity.

$$Total\ Delay = \ Equivalent\ in\ Delay\ \times\ Mean\ number\ of\ passengers\ per\ cancelled\ flight$$

$$Cost\ per\ Day = \ Total\ Delay\ \times\ Recommended\ Hourly\ Value\ of\ Travel\ Time\ savings$$

**Passengers standpoint** The passenger cost of having his or her flight cancelled can be estimated from aggregated statistics. In 2007, 11.4 million passengers were affected by flight cancellations, for a reported cost of 3.2 million $ [36].

$$Passenger\ cost\ per\ cancellation\ = \frac{Passenger\ cost\ in\ 2007}{Number\ of\ passengers\ disrupted\ due\ to\ cancellations} = 281.5\ USD$$

Table 5 summarizes the passenger costs for departure cancellations at SFO airport. The same computation is performed for arrival cancellations.

### 3.3.3 Diversions

**Airplane stand point** Diversions are an expensive, chronic and disruptive element of flight operations, costing at least 300 million dollars annually to US carriers for domestic flights alone. A diversion is not a single, discrete event, but rather a set of cascading actions that cause severe disruptions to airline schedules, major costs, and significant passenger frustration. Diversion costs from the airplane standpoint can range from 15,000 $ for a narrow-body domestic flight, to more than 100,000 $ for a wide-body international flight. Perry Flint, from the International Air Transport Association (IATA) [37], stated that there is no average cost for diversions. The diversion cost depends on the size of the aircraft and the number of passengers on board, severity and operational consequences: delays, resulting missed connections, flight cancellations, cost of hotels/meals for passengers and crew.

Table 5: Cost of cancellations during the crash week-end

| | Cost of cancellations (million $) |
|---|---|
| Cancellations from SFO (Airplane) | 6.2 |
| Cancellations from SFO (Passengers) | 12.5 |
| Cancellations in SFO (Airplane) | 5.1 |
| Cancellations in SFO (Passengers) | 10.2 |
| Total Cost | 34.0 |

Assume that the diversion cost depends on the diversion airport. If the diversion airports are OAK or SJC, the total diversion cost from the airplane standpoint can be assumed to be negligible. From Saturday, July 6th, to Tuesday, July 9th, 40 flights were diverted to SJC and 39 to OAK, as pointed out in the previous diversion analysis. For all the other airports accommodating domestic diverted flights (LAS, LAX, DEN, MSP, PHX, RNO, SLC, SMF), the average diversion cost is chosen to be 25,000 $. Moreover, flights that finally arrived in SFO are distinguished from flights that stayed at the diversion airports. For the flights that finally arrived in SFO, the cost is set to 0 dollars since the diversion cost for the passenger standpoint is accounted for through recorded flight delays. There were 44 diversions in SFO during the crash week-end. Each flight that stayed in the diverted airports, incurred a cost of 25,000 $.

**Passengers stand point** For the diverted flights that finally reached SFO, their arrival delay is computed and the same value of hourly travel time is used to estimate the cost from the passengers point of view.

$$Total\ Delay = \sum_{\substack{i=0 \\ i}}^{n} Aircraft\ Delays_i \times Passengers\ per\ Aircraft\ i$$

The overall cost of diversions adds up to 0.9 million $.

For the flights that stopped at their diversion airport, a few assumptions are made :

- All the passengers had to go to SFO, even if they landed in OAK or SJC.
- Once a passenger is diverted to another airport and lands, he/she has to take a car or train to go back to his/her original destination, SFO. For that segment of the journey, the multimodal passenger speed is assumed to be 60 mph.
- If the distance between SFO and the diverted airport is greater or equal to 600 miles, the passengers is unlikely immediately take a car to drive to SFO by him or herself but rather wait to be reaccommodated by the airline. In this case, the flight cost is considered to be the same as the cost of a cancellation.

Under these assumptions, the five domestic flights that landed in DEN, HNL, MSP and PHX are considered to be equivalent to cancelled flights to estimate the cost. On board these flights were 728 passengers, and the associated approximate cost is therefore $Cost\ of\ diverted\ flights\ (distance > 600\ miles) = \$\ 205,000$.

Regarding the other flights diverted to airports less than 600 miles away from SFO airport, the total distance adds up to 8 million miles. Using the hypothesis that a passenger speed is 60 mph, the total delay to SFO is 133,633 hours. The associated cost is 3,8 million $. Combining the three sources of cost, the total diversions cost for the passengers point reaches 4,9 million $.

Using the previous method and summing all delays, the total flight delay during the crash week-end is equal to 27,067 hours and the associated cost is 774,125 $.

**Diversions cost** The total cost of diversions during the crash week-end is thus estimated to reach 32.3 million dollars.

Table 6: Costs of diversions during the crash week-end

| | Cost of diversions (million $) |
|---|---|
| Diversions of scheduled departures from SFO (Airplane) | 0 |
| Diversions of scheduled departures from SFO (Passengers) | 0.75 |
| Diversions of scheduled arrivals to SFO (Airplane) | 0.6 |
| Diversions of scheduled arrivals to SFO (Passengers) | 4.9 |
| Total Cost | 5.95 |

### 3.3.4 Overall Cost

The overall cost of the crash is computed as the difference between the cost of the crash week-end and that of the previous "normal" week-end. The cancellations costs due to the crash is estimated at 31.37 million dollars, or 64% of the total crash cost. The delay cost is estimated at 11.6 million dollars, or 24% of the total cost. The diversions cost is the lowest , accounting for 12% of the total crash cost. Thanks to these observations, it is clear that the important cost of the crash is mainly due to the cancellations and the important delays during the week-end.

It should be noted that the overall cost calculated does not take into account the cost of passenger losses, passenger evacuation and hospitalization, emergency interventions on the crash scene, airport repairs and aircraft loss.

| Cost (Million dollars) | Crash week-end | Week-end before | Difference |
|---|---|---|---|
| Delay in SFO (Passengers) | 8.0 | 0.4 | 7.6 |
| Delay in SFO (Airplane) | 6.1 | 0.3 | 5.8 |
| Delay from SFO (Passengers) | 5.6 | 1.6 | 4.0 |
| Delay from SFO (Airplane) | 3.9 | 1.3 | 2.6 |
| Cancellations in SFO (Airplane) | 5.1 | 0.3 | 4.8 |
| Cancellations in SFO (Passengers) | 10.2 | 0.4 | 9.8 |
| Cancellations from SFO (Airplane) | 6.2 | 0.2 | 6.0 |
| Cancellations from SFO (Passengers) | 12.5 | 0.3 | 12.3 |
| Diversions in SFO (Airplane) | 0.6 | 0.0 | 0.6 |
| Diversions in SFO (Passengers) | 4.9 | 0.0 | 4.9 |
| Diversions from SFO (Airplane) | 0.0 | 0.1 | -0.1 |
| Diversions from SFO (Passengers) | 0.8 | 0.1 | 0.7 |
| **Total** | **63.9** | **4.9** | **59.0** |

Table 7: Cost evaluation for the crash week-end.

This cost decomposition analysis suggests that the costs incurred to the air transportation system could have been strongly reduced if several delayed and/or cancelled flights arriving in SFO had been diverted to OAK or SJC, if additional gate and runway capacity were available.

# 4 Impact of the Crash on other Transportation Modes

## 4.1 Ground Transportation : Highway Traffic

The major data source for the road network in California is PeMS, which stands for Freeway Performance Measurement System [38]. Measurements from loop detectors on the major roads in California are recorded and stored in PeMS. The loop detectors measure the number of vehicles passing per time period (flow) and the fraction of time that the loop is occupied (occupancy). From these measurements, a number of traffic properties are estimated, for example the average speed of vehicles on a given road. However, the PeMS data presents two main limitations. First the traffic conditions on a road stretch between two detectors are not observed. Second, many loop detectors are out of order for periods of time. The second limitation has an compounding effect on the first limitation.

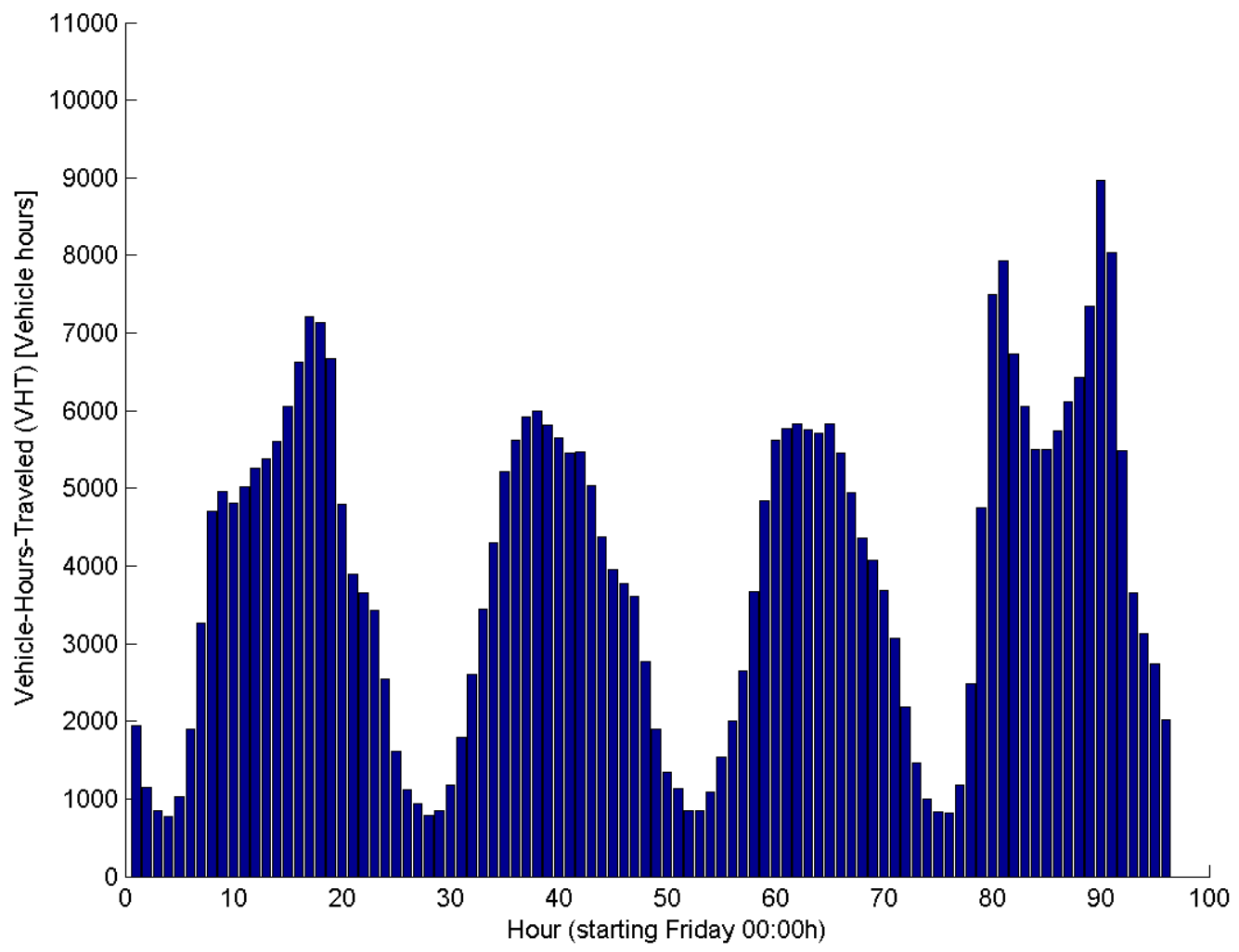

(a) Vehicle Hours Traveled

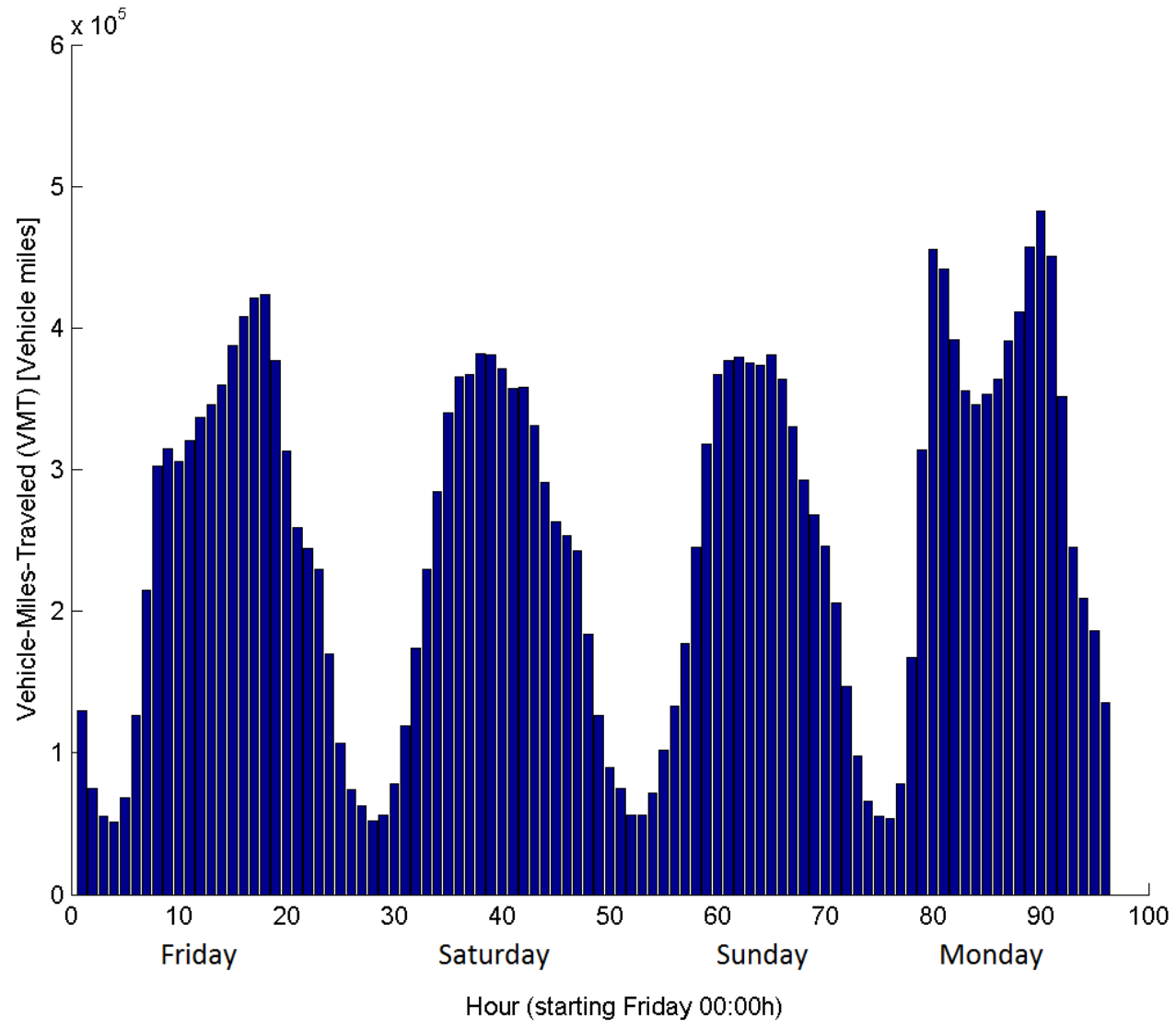

(b) Vehicle Miles Traveled

Figure 16: Road traffic performance data for the long weekend in which the ASIANA crash occurred.

Choe [39] proposes a method to analyze road traffic conditions using PeMS. PeMS has been used in several studies to study congestion growth [40]. For the present study, the hourly Vehicle-Hours-Traveled (VHT) and Vehicle-Miles-Traveled (VMT) in an eight-mile radius [1] around SFO airport are studied from Friday, July 5th to Monday, July 8th, as displayed in Figure 16.

The variables used to understand road traffic are defined as follows. $q$ is the flow in [veh/h], $T$ is the time period in [h], $L_s$ is the length of the considered road stretch in [mi], and $v$ and $v_r$ are the observed and reference speeds in [mph], respectively.

[1] This data is not a direct output of PeMS. The VHT and VMT for the different PeMS roads stretches within this area are summed.

$$Delay = qTL_s\left(\frac{1}{v}-\frac{1}{v_r}\right) \tag{1}$$

$$VMT = qTL_s \tag{2}$$

$$VHT = qT\frac{L_s}{v} = kTL_s \tag{3}$$

$$Speed_{avg} = \frac{VMT}{VHT} \tag{4}$$

To better understand performance data based on certain traffic conditions, consider an hypothetical road stretch of one mile during a period of one hour. For this space and time, assume uniform traffic conditions. Table 8 shows the resulting performance variables. This example only works for uniform traffic conditions, but gives a good indication of the contribution of traffic conditions types.

Table 8: Performance variables for three one-mile road stretches with uniform traffic conditions during one hour

| State | Flow [veh/h] | Speed [mph] | Delay[1] [veh h] | VMT [veh mi] | VHT [veh h] |
|---|---|---|---|---|---|
| 1 | 6000 | 65 | 0 | 6000 | 92.31 |
| 2 | 7900 | 64 | 1.90 | 7900 | 123.44 |
| 3 | 7180 | 50 | 33.14 | 7180 | 143.60 |

[1] Reference speed is 65 mph.

A traffic state is defined for a given flow rate and speed, for a specified time period and road stretch. State 3 in Table 8 is the only congested state. In the road stretches where this state is observed, the average speeds are lower and the delays higher than in the other road stretches. From this example, congestion leads to a large delay and low speed. Near capacity (synchronized) flow contributes mostly to VMT.

Table 9: Performance variables for six road stretches consisting of two one-mile road stretches with uniform traffic conditions during one hour

| States | $Speed_{avg}$ [mph] | Delay[1] [veh h] | VMT [veh mi] | VHT [veh h] |
|---|---|---|---|---|
| 11 | 65 | 0 | 12000 | 184.62 |
| 22 | 64 | 3.80 | 15800 | 246.88 |
| 33 | 50 | 66.28 | 14360 | 287.20 |
| 12 | 64.43 | 1.90 | 13900 | 215.75 |
| 13 | 55.87 | 33.14 | 13180 | 235.91 |
| 23 | 56.47 | 35.04 | 15080 | 267.04 |

[1] Reference speed is 65 mph.

Now consider a two mile road stretch during a one-hour period, with two equal length uniform traffic conditions. The corresponding performance data is depicted in Table 9. The VMT and VHT are both larger in state 22 than in state 13. It is not always the case when there is a combination of free-flow and congestion, yet it indicates that it is possible. It corresponds to light congestion and free-flow that is not close to capacity.

The traffic performance data around SFO displayed on Figure 16 do not indicate any clear effect of the crash on the aggregated traffic conditions for Saturday, July 6th. However, a more detailed analysis can be performed using the space-time contour plots described in [39]. We define "abnormal congestion" as congestion that does not occur on reference days and thus is not caused by regularly occurring bottlenecks. The road traffic conditions are compared with reference days, namely with the 26 Saturdays between April and September 2013. Using the method in described in [39], the congestion on on US101, I80 and I880 near SFO is recorded for each of the Saturdays. The congestion on the US101N near SFO stands out as abnormal. At all other locations and times where congestion was observed, congestion had also occurred at least once during the reference Saturdays. Therefore, the rest of this subsection focuses on the US101N to observe the traffic jam reported in the news [41]. When it comes to freeway traffic, under nominal conditions, week days exhibit morning and evening peaks, but week ends do not. Figure 16 shows that only the congestion pattern on Friday, July 5th is different. This is expected for a normal long-weekend, with no morning peak because of July 4th on Thursday. A pattern change on Saturday due to the ASIANA crash is not clear from these figures because the data is too aggregated.

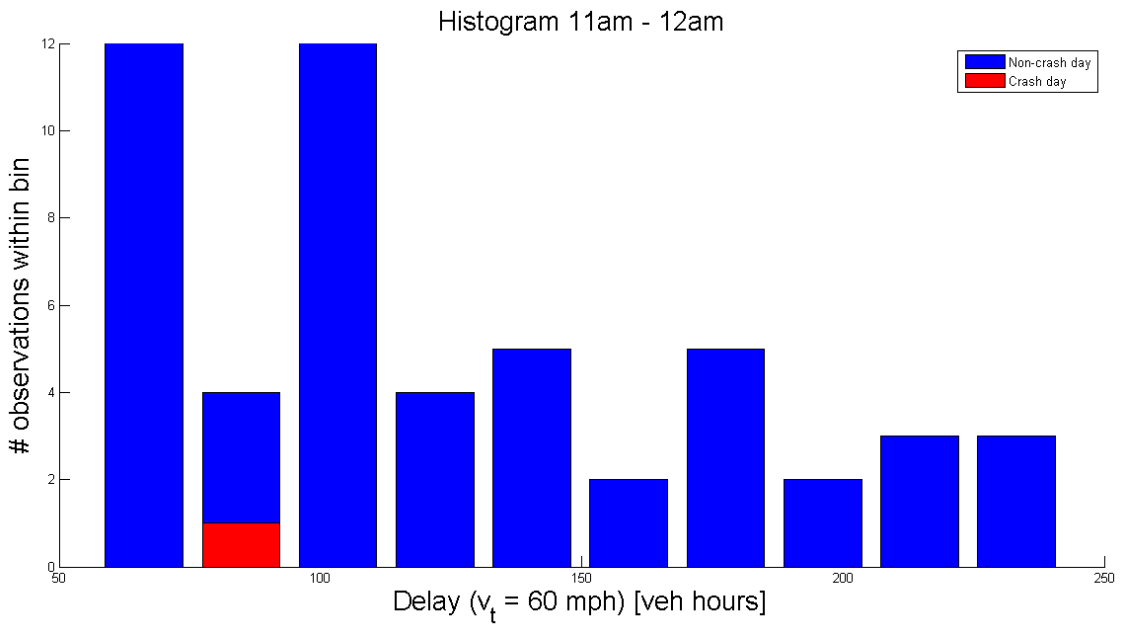

(a) Delay 11am-12am

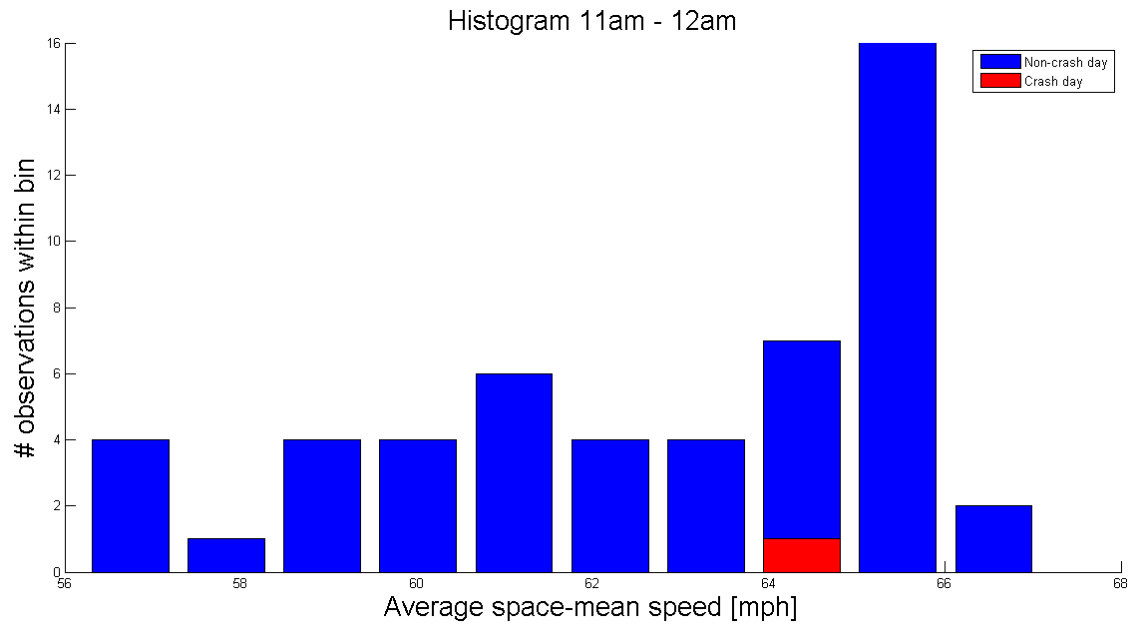

(b) Speed 11am-12am

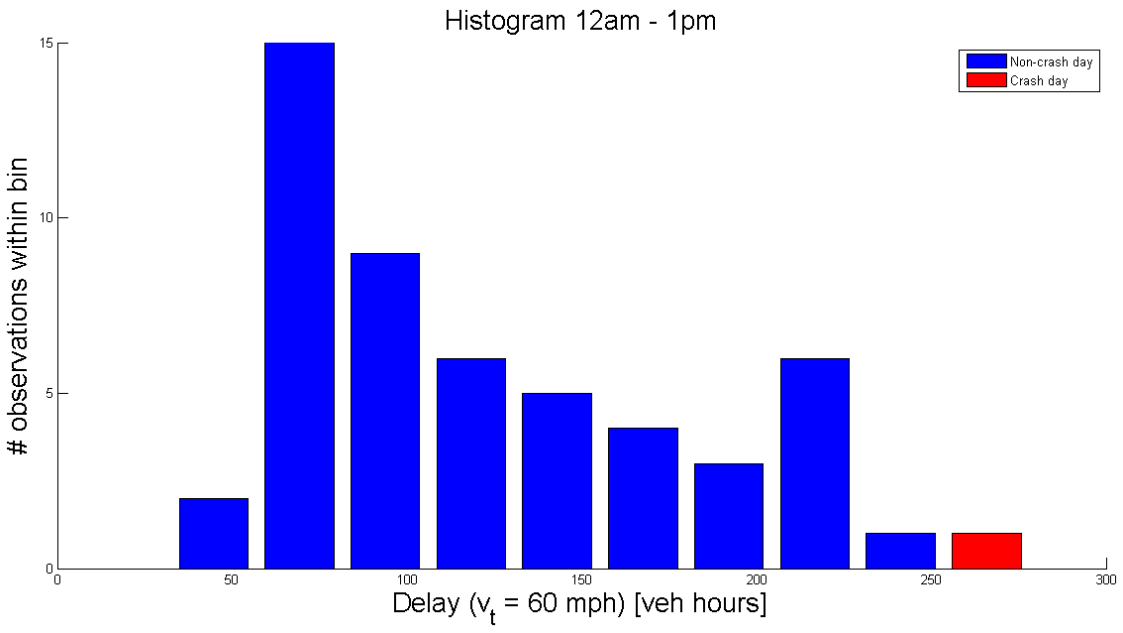

(c) Delay 12am-1pm

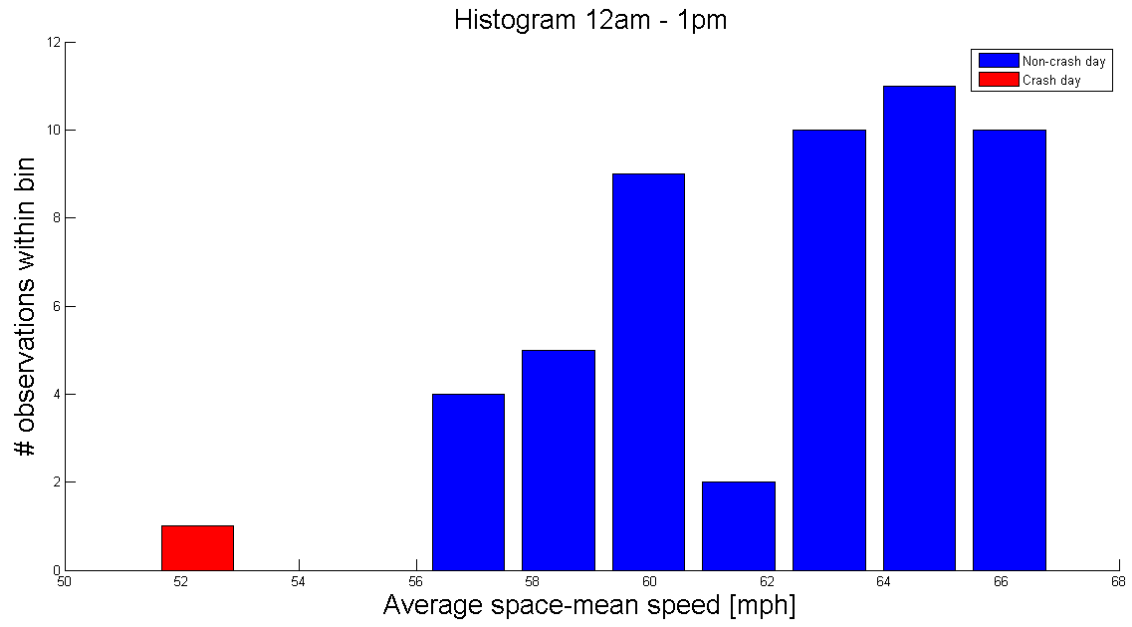

(d) Speed 12am-1pm

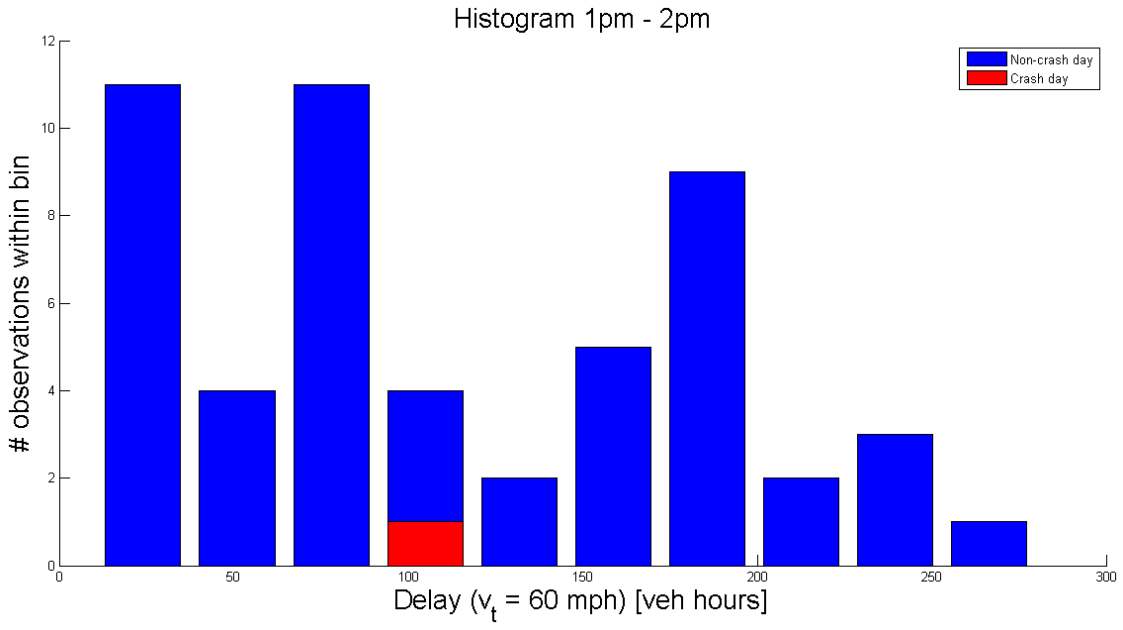

(e) Delay 1pm-2pm

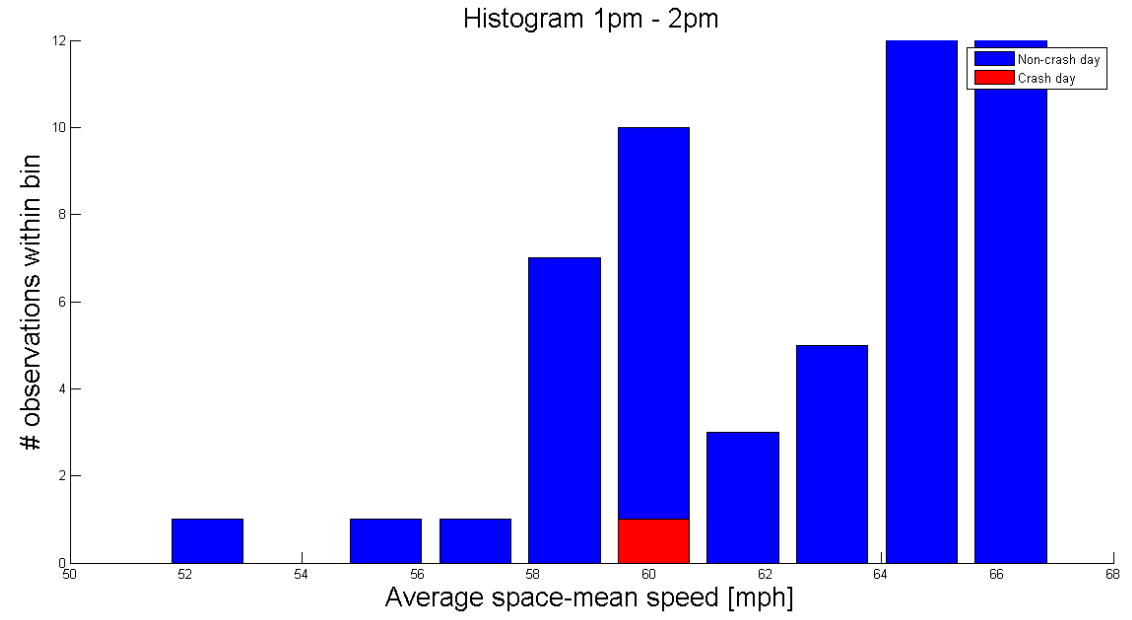

(f) Speed 1pm-2pm

Figure 17: Histograms of the delay and speed on the US101-N PM 415-425 for the 26 Saturdays between April and September 2013.

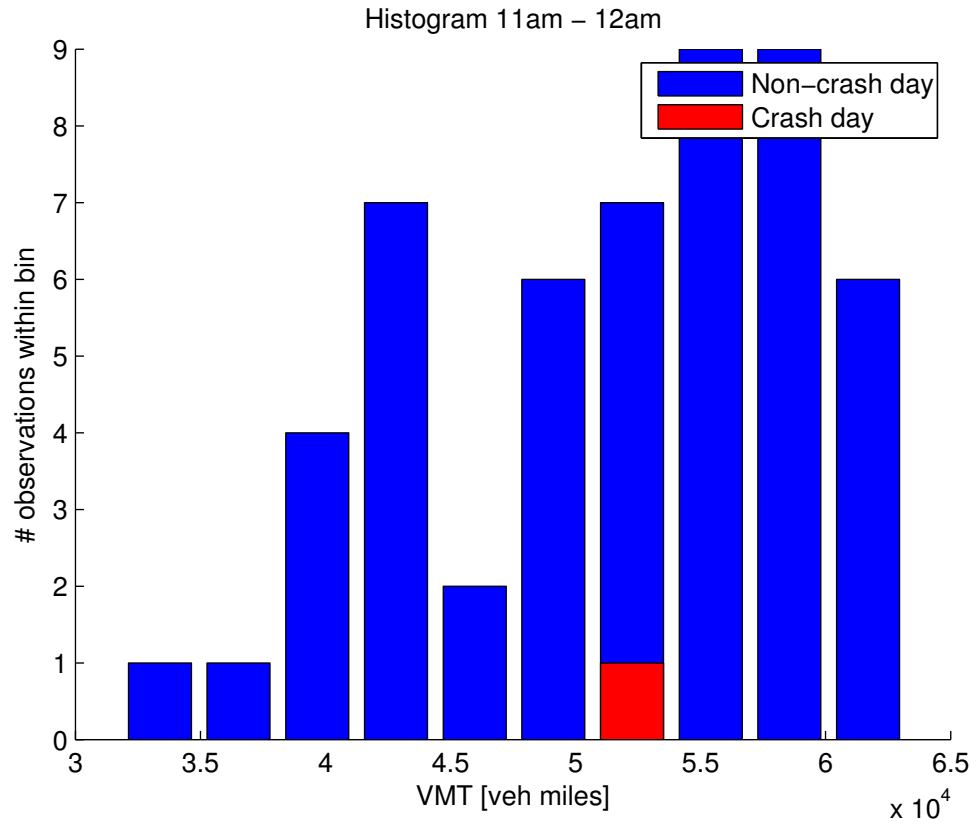

(a) VMT 11am-12am

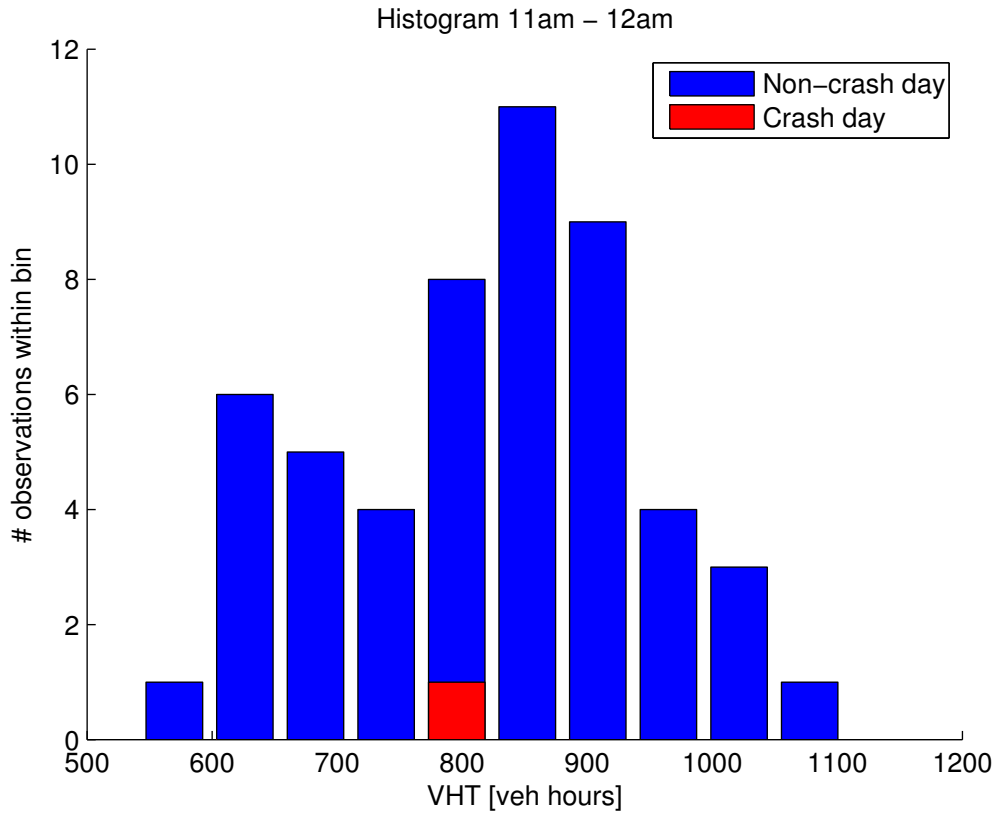

(b) VHT 11am-12am

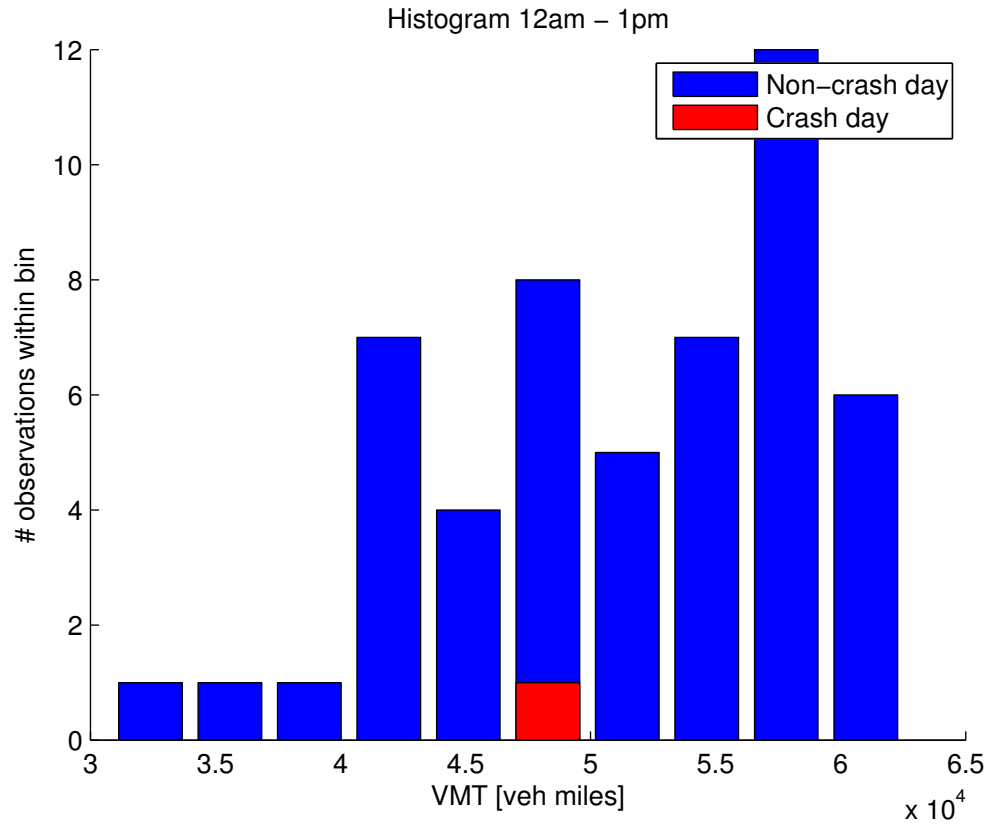

(c) VMT 12am-1pm

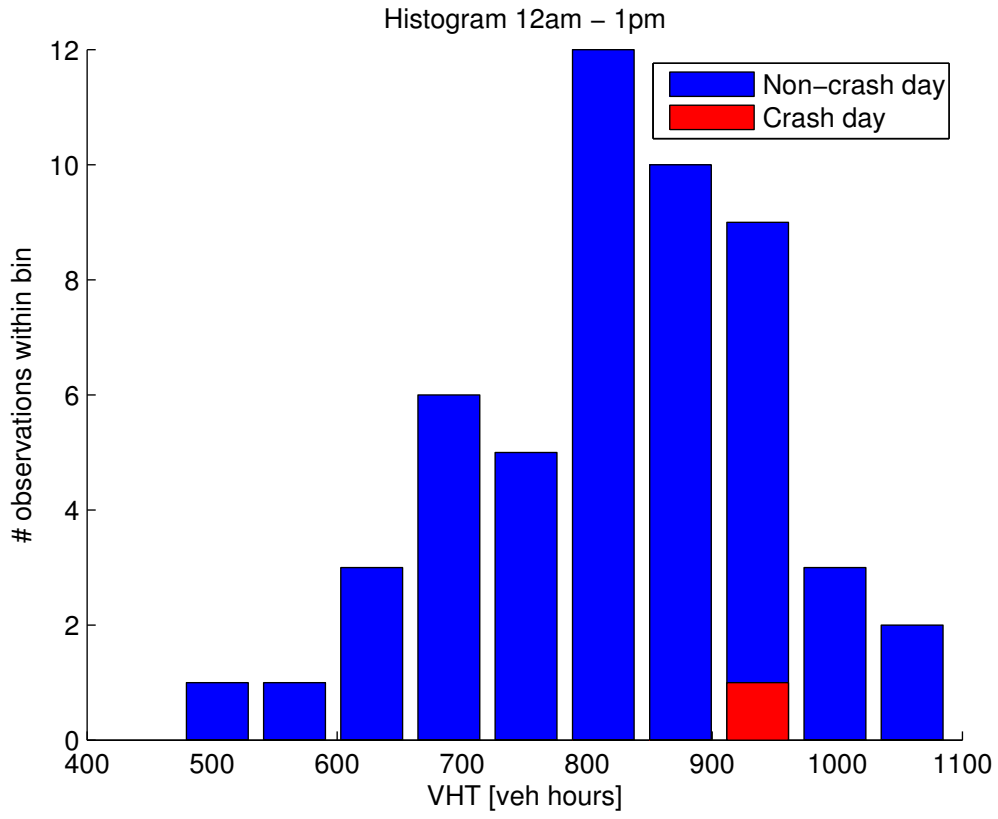

(d) VHT 12am-1pm

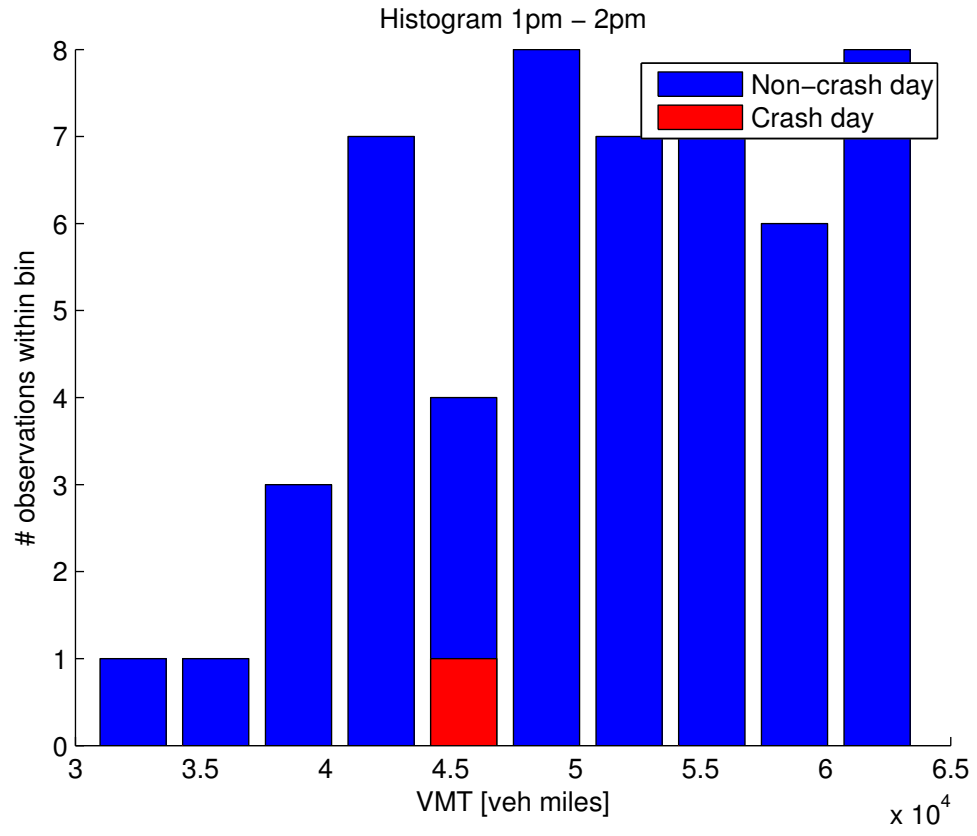

(e) VMT 1pm-2pm

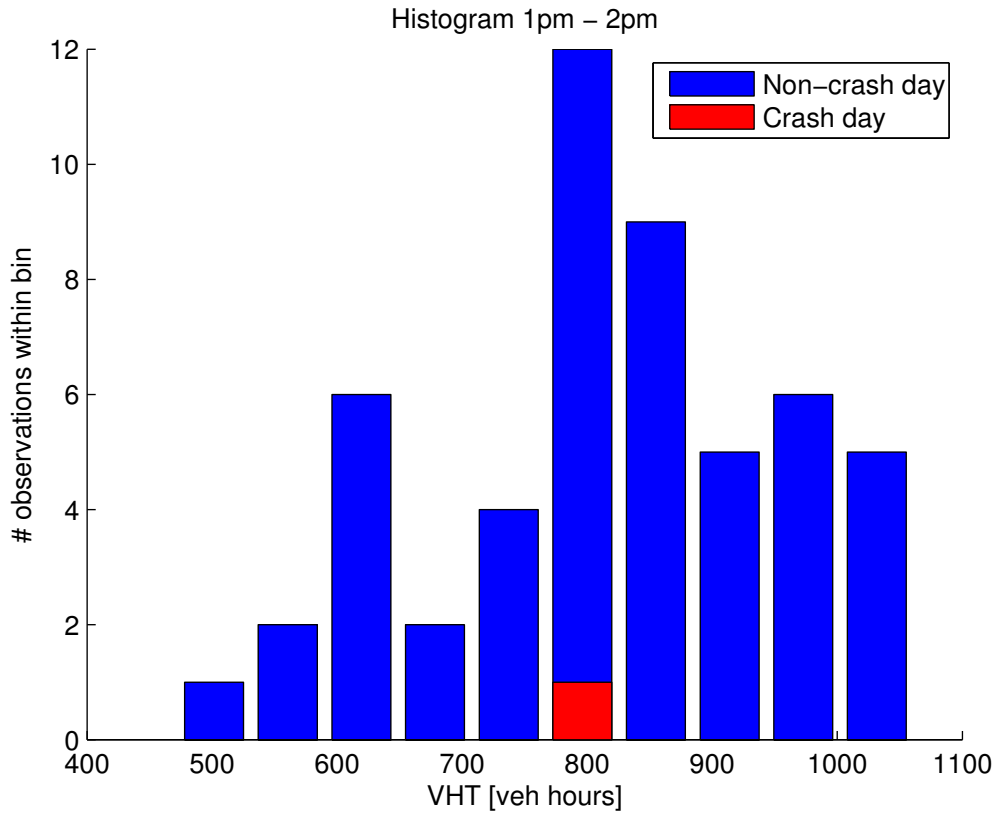

(f) VHT 1pm-2pm

Figure 18: Histograms of the VMT and VHT on the US101-N PM 415-425 for the 26 Saturdays between April and September 2013.

The performance data (speed, delay, VMT, VHT) at a US101N road stretch near SFO are shown in Figures 17 and 18. From the speed and delay histograms in the 12 am - 1 pm period, the speed and delay on the crash day stand out as outliers. On the other Saturdays in 2013, the delay was never as large and the speed never as low as on the ASIANA crash day. From the VMT and VHT plots in Figure 18, no clear effect of the ASIANA crash is observed after 1 pm on the crash day. The breakdown occurred shortly after the ASIANA crash and directly next to SFO. The present focus is on understanding the causes of the congestion and, if possible, understand the causality is between the crash and congestion. Therefore, an in-depth analysis is performed into this congestion.

As a starting point, the following hypotheses about the potential causes of the congestion are formulated:

- Increase in demand caused by vehicles leaving the airport.
- Rubbernecking, a traffic breakdown is caused by users watching the crash and thereby changing their behavior.
- Effect caused by emergency vehicles trying to reach the airport.
- Accident on the highway.
- Lane or ramp closure.

In order to investigate whether external events other than the ASIANA crash, such as an accident, caused the congestion, we consider the California Highway Patrol (CHP) incidents feed on the US101N near SFO on the crash day. The CHP incidents feed provide the incidents reported on that specific road stretch and time period. The Abs PM correspond to the first off-ramps upstream and downstream of SFO, it indicates the location of the road stretch studied. Here all reported types, namely accident, hazard, breakdown, police, congestion, weather and other are included. Table 10 shows the CHP incidents reported on July 6th, 2013 on the US101N near SFO. The CHP incidents feed provides no indication for the cause behind the observed aberrant congestion. All reported incidents occurred either after the aberrant congestion and/or on different locations. PeMS also has a lane closure system, in which the historical lane closures are reported. On the ASIANA crash day, two lane closures were reported by the system, see Table 11. We assume that these combined indicate that the off-ramp to SFO was closed at 12:30pm.

Table 10: CHP incidents feed on the US101N between PM 400-430 on July 6th, 2013 between 8am and 10pm

| Time | Dur. [min] | Abs PM | Location | Description |
|---|---|---|---|---|
| 15:26 | 18 | 425,5 | San Francisco Us101 N / Us101 N Sierra Point Pkwy Ofr | 1125-Traffic Hazard |
| 12:18 | 27 | 404 | Redwood City Us101 N / Willow | 20002-Hit and Run No Injuries |
| 12:41 | 19 | 414,7 | Redwood City Us101 N / Us101 N Kehoe Ave Ofr | 1179-Traffic Collision-1141 En route |
| 19:25 | 6 | 418,5 | Redwood City Us101 N / Us101 N Broadway Ofr | 1125-Traffic Hazard |

None of the above CHP notations are related to the breakdown near SFO airport.

Table 11: Lane closures on the US101N between PM 400-430 on July 6th, 2013 between 8am and 10pm

| Begin | End | Abs PM | Facility | Closure Lanes |
|---|---|---|---|---|
| 12:30 | 16:30 | 420,134 | Off Ramp | 2 |
| 12:30 | 16:30 | 422,572 | Off Ramp | 2 |

The Abs PM correspond to the first off-ramps upstream and downstream of SFO. This can either mean that the two off-ramps were closed or that the one in-between (the SFO off-ramp) was closed. With additional information from news reports and tweets, the most likely meaning is that the off-ramp to SFO was closed.
However it is unclear whether the off-ramps were closed for the entire period. The last update was at 13:54, which could correspond to the reopening of the off-ramps. The timing could also match an announcement of the reopening of the two shorter runways at SFO. Because the congestion observed is mostly between noon and 1 pm, therefore before 13:54.

To study the congestion over time, the animation function available in PeMS [38] is edited to highlight the important moments. In Figure 19, screen shots of the animation tool are displayed to highlight the main events on the highway. At 11:26am, the ASIANA aircraft crashed just short of one of the SFO runways, see Figure 19a. At 11:53am, the first breakdown occurred at the on-ramp from Millbrae (PM 420.5), see Figure 19b, just upstream of the off-ramp to SFO. At the same time, conditions (lower speed) are deteriorating at the second upstream on-ramp (PM 419). However, a further breakdown first occurs at PM 420.5. Later, around 12:06pm, a further breakdown happens at PM 419, see Figure 19c. This creates the heavy congestion between PM 417-419, as previously observed. Around 12:49 pm the largest road stretch of the US101N is congested, as seen in Figure 19d with the red color. The congestion does not dissolve until 1:30pm, then the traffic conditions are restored to normal. In Figure 19e some lower speeds are spotted on the PM 419 on-ramp.

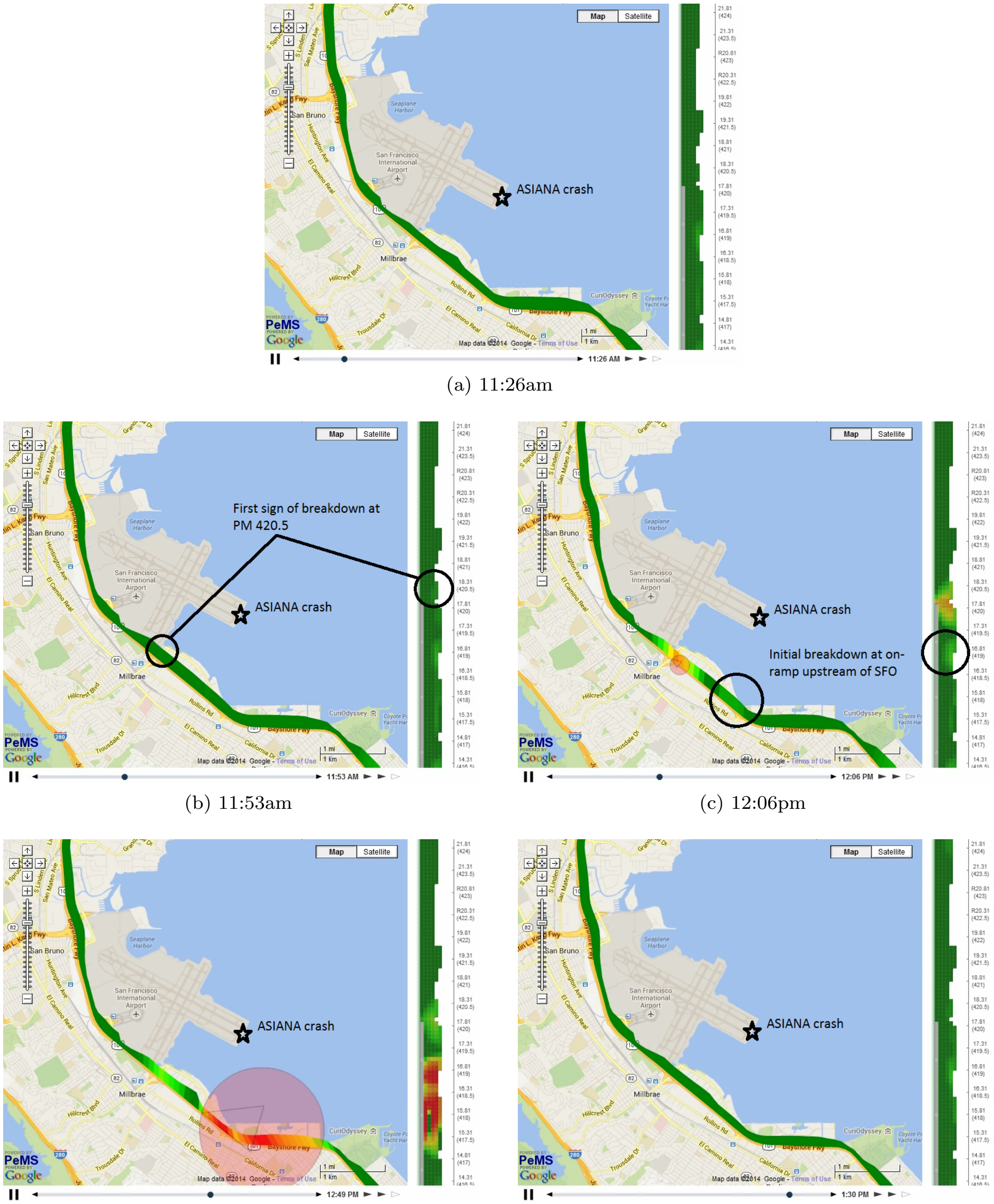

(a) 11:26am

(b) 11:53am

(c) 12:06pm

(d) 12:49pm

(e) 1:30pm

Figure 19: Traffic situation on the US101N at different times on July 6th, 2013. Two breakdowns can be observed: the first at 11:53 am, the second at 12:10 pm.

These observations suggest that the congestion was not likely to be caused by a large number of road vehicle departures from SFO (or inflow on the US101N) after the crash, because congestion occurs far upstream of the SFO to US101N on-ramp. The congestion observed on the US101N road stretches between PM 416.3 and 420.9. This range was selected such that the most upstream and downstream detectors show no sign of congestion. For further analysis, the road lay-out and individual loop detector stations on the mainline and ramps of the important road stretch are considered, as shown in Figure 20.

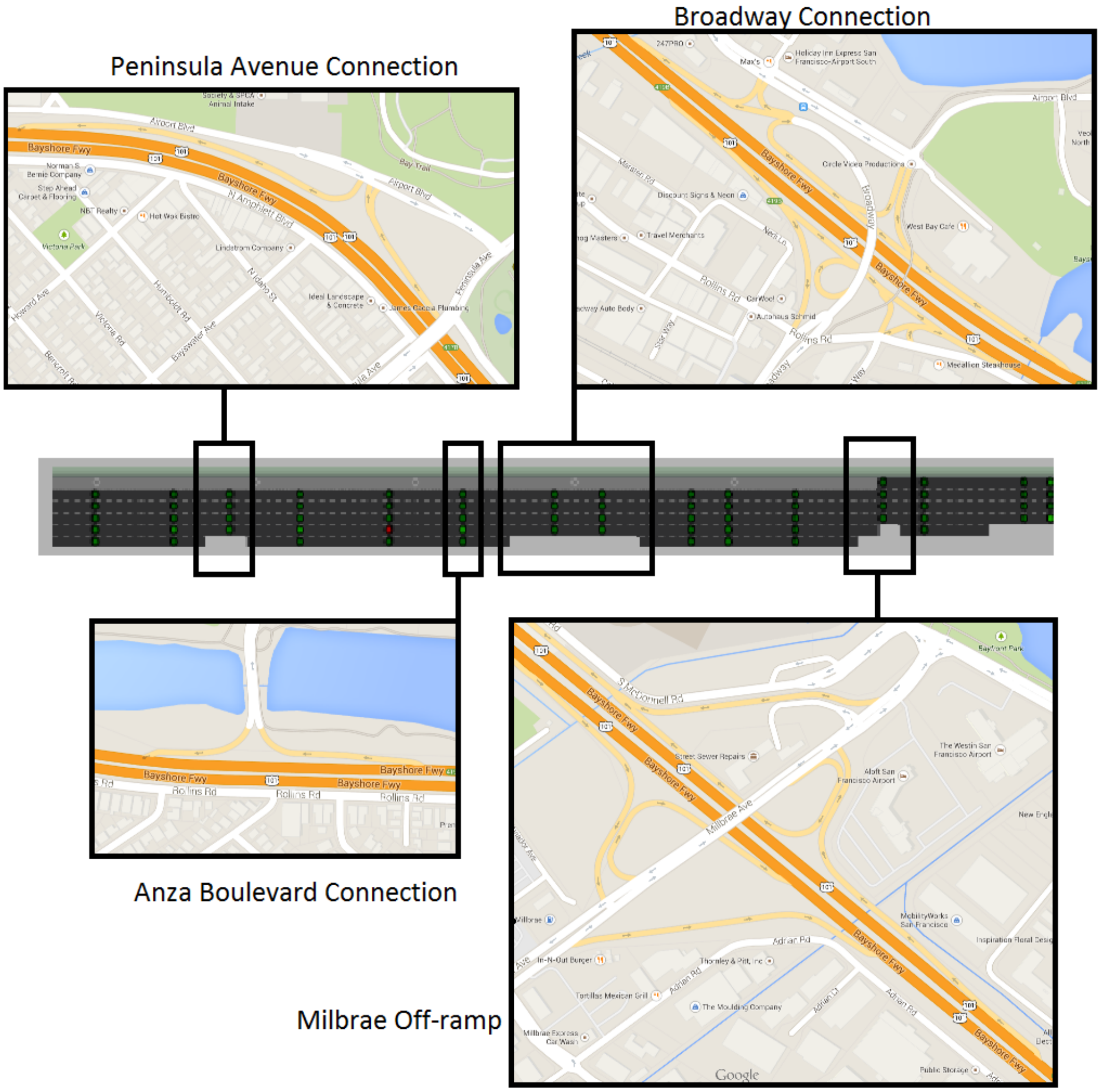

Figure 20: Lay-out of the US101N road stretch where congestion occurs near SFO after the ASIANA crash. The three connections, namely Peninsula Avenue, Anza Boulevard and Broadway, consist of both off and on-ramps, while the traffic coming from the Millbrae off-ramp stays on a secondary road until after the considered road stretch.

For the considered road stretch, fifteen mainline stations were available. Their locations are shown in Table 12. Besides these stations, traffic information is available on the Anza Boulevard on- and off-ramp and one of the two Broadway on-ramps, namely about the vehicles using the fly-over.

For all stations, the occupancy and flow is decomposed in 5-minute time-periods. For these time-periods, the average space-mean speed is also available at the mainline stations. Following [39], the present focus is on the occupancy over time on the different station locations. The first station is located downstream and the last upstream of the congestion on US101-N. At this location no congestion occurs, see Figure 21a. The occupancy upstream of the congestion remains stable between 0.070 and 0.075 (fraction of time a detector is occupied), while it varies more at the downstream station. However, the occupancy there remains under 0.100, indicating that there is no or very limited congestion. Between 12:00 pm and 12:30 pm, a break occurs, causing the occupancy to drop and oscillate around 0.065. This indicates that after this period, fewer vehicles

Table 12: The fifteen considered detector stations on mainline US101N

| Detector station number | Abs PM | Detector station number | Abs PM | Detector station number | Abs PM |
|---|---|---|---|---|---|
| 1 | 420.887 | 6 | 419.407 | 11 | 417.847 |
| 2 | 420.767 | 7 | 419.237 | 12 | 417.437 |
| 3 | 420.307 | 8 | 418.827 | 13 | 417.117 |
| 4 | 420.117 | 9 | 418.607 | 14 | 416.857 |
| 5 | 419.717 | 10 | 418.187 | 15 | 416.497 |

The Abs PM correspond to the locations of first off-ramps upstream and downstream of SFO.

use the US101-N at this location. At that point in time, the congestion may have been known to users and shortly after the authorities asked people to use the I280 instead of the US101. The fact that congestion clear afterwards may suggest that people listened to the calls of the authorities.

A decomposition of the congestion pattern on US101 is provided in Figure 19, showing that two breakdowns occur. The first breakdown happens close to the Millbrae connection. The occupancy measured at the three detectors in the affected road stretch is shown in Figure 21b. The first peak indicates the breakdown timed at 11:53 am with the video-animation. Yet, this does not provide new insight regarding the potential causes of this breakdown.

The second, more severe, breakdown occurs at the Broadway connection, where the causes may be better understood: the congestion resulting from the second breakdown is clearly observed by the seven detector stations shown in Figure 21c. This Figure shows that the congestion starts at the most downstream stations, as the PM 419.237 and PM 418.827 station first show a higher occupancy. A jump is observed during the 12:10-12:15 time-period. Although a decrease in speed is noted just before that in the video-animation, the breakdown in this period is displayed in 21c.

At the Broadway connection, only the detectors on the on-fly US101N on-ramp are working. Figure 21d shows the flow and occupancy over time on this on-ramp. A large, single-period peak in the flow is observed. This peak occurs within the same time-period as the breakdown, between 12:10-12:15. The increase in on-flow at Broadway thus coincides with the breakdown in the same location. It is the most likely cause of the breakdown. Although congestion may not have been purely caused by this increase in demand, it is at least one of the most contributing factors.

One probable explanation is that the congestion corresponds to 'extra' emergency vehicles dispatched to assist with the Asiana crash. From the Asiana crash investigation report, a transcript states that : "By 11:33,(...) all seven airport firefighting companies and paramedics were on scene. (...) One minute later, 56 ground ambulances arrived on scene. (...) At 13:01, the last patient was transported by ambulance." [42]. Additionnally, a helicopter and two buses also helped transport patients to 12 area hospitals.

The ASIANA crash affected the ground traffic conditions. Although there were multiple breakdowns, we can only state that the congestion on the US101N near SFO was a direct consequence of the crash. The (visible) smoke, the disturbance due to emergency vehicles and the ramp closures are the most likely causes. However, further details about the coupling between the airside and the highway system remains difficult to quantify.

## 4.2 Ground Transportation : Public Transit with the BART

The BART, or Bay Area Rapid Transit, is one key element of the transit transport in the San Francisco Bay. It links SFO to OAK as well as SJC via the Caltrain connection, see Figure 22. The BART data obtained provides the origin-destination matrix of passengers for 15 minutes periods on Saturday, June 29th and Saturday, July 6th.

The comparison between June 29th and July 6th for departing and arriving passengers at the SFO BART station, displayed in Figure 23 shows that the total number of passengers using this transit station was smaller on the crash day, with up to 100 fewer passengers at peak hours. Because the airport was closed for part of the day and many flights were cancelled, we can hypothesize that simply fewer passengers used this transit station.

Next, passenger traffic at the OAK BART station is studied. The results are displayed in Figure 24 for passenger traffic between SFO and OAK, between OAK airport and SFO airport. In both directions between the two airports, there is a significant increase of passengers soon after the crash. Between SFO to OAK, there is very little traffic on both Saturday, June 29th and Saturday, July 6th, with fewer than 10 passengers

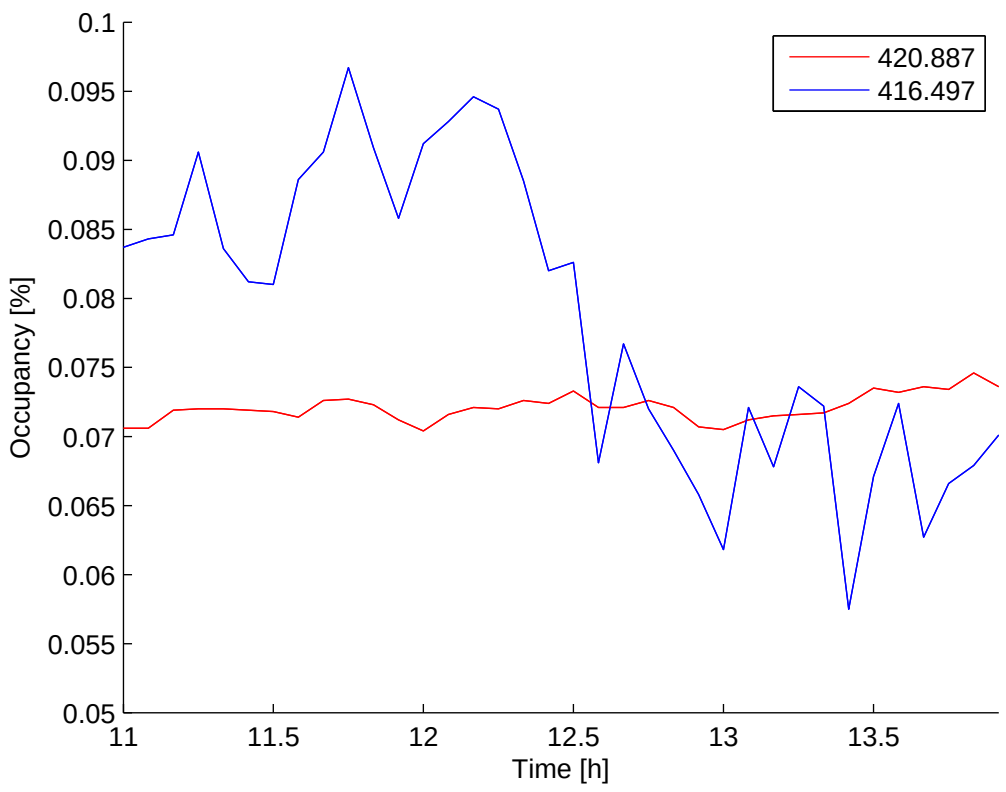

(a) Occupancy over time for the first (downstream) and fifteenth (upstream) considered stations

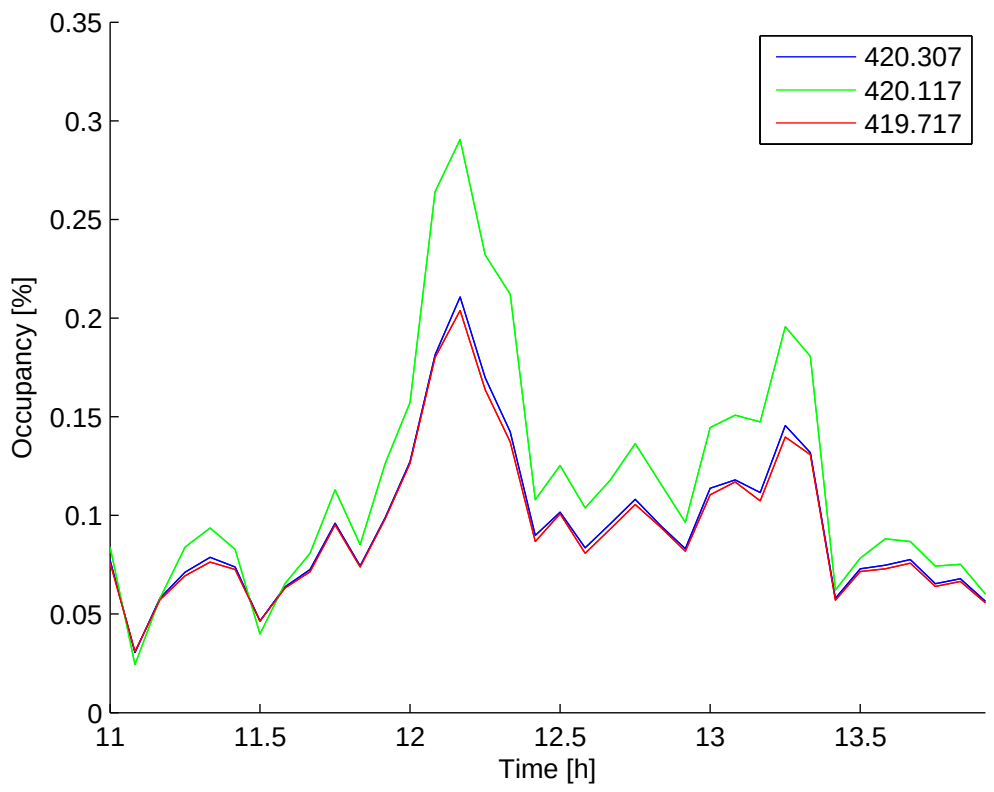

(b) Occupancy over time for the first breakdown

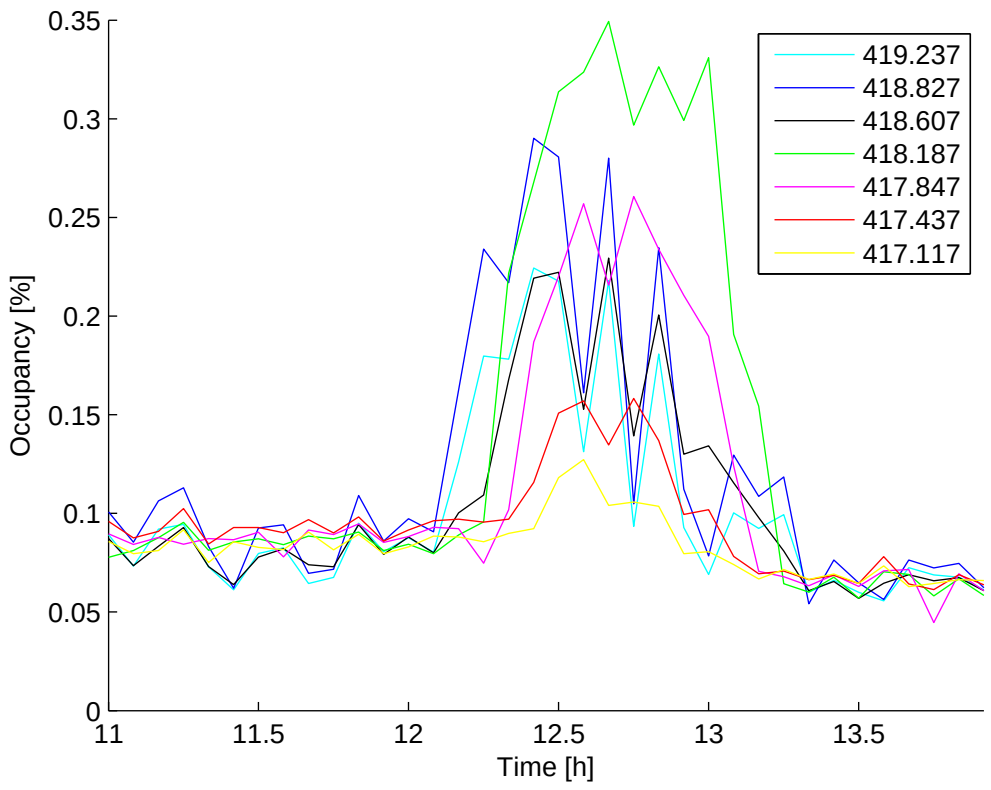

(c) Occupancy over time for the second breakdown

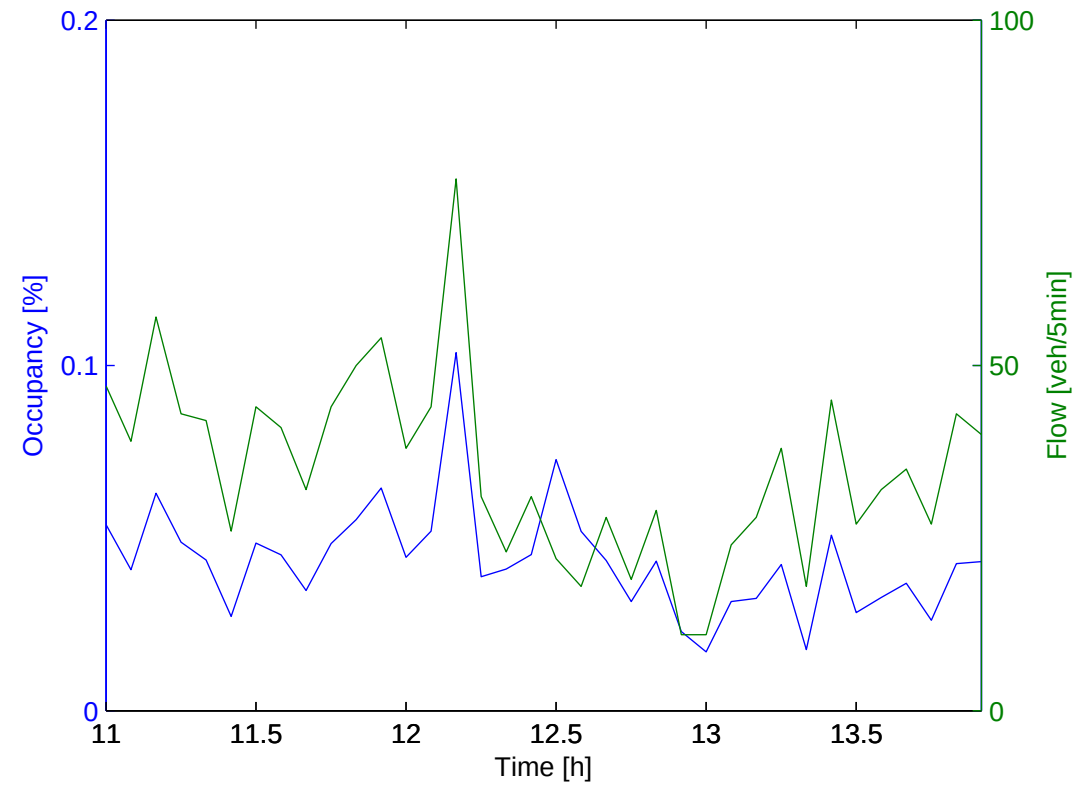

(d) Occupancy and flow over time on the Broadway on-ramp to the US101N.

Figure 21: Occupancy on US101 highlighting the two breakdowns after the crash.

per hour. On the crash day, between 2 pm and 3 pm, up to 160 passengers choose to travel from SFO to OAK. The reason behind this is still unclear : these passengers could be trying to reach air travellers diverted to OAK airport, or airline employees could be suddenly needed to accommodate the incoming air traffic at the airport. The Oakland to San Francisco passenger traffic is also an outlier on the crash day, but the number of additional passengers is less noteworthy. Both abnormal patterns persist throughout the day.

# 5 Conclusion

A case report of the Asiana crash in San Francisco International Airport on July 6th 2013 and its repercussions on the multimodal transportation network is proposed.

Transportation networks are intrinsically tied or coupled. In the present study, we consider the air, road and rail transit networks. It must be noted that these networks exhibit interdependencies with other networks, such as the power and communication networks for instance. Networks are usually studied separately. To the best of the authors' knowledge, this paper constitutes the first study of interdependencies between transportation networks. When studied individually, networks may appear to have a fairly robust structure towards random failures. However, when their coupling with other networks is taken into account, their sensitivity is higher than when studied independently. The ASIANA crash is a powerful example of node failure leading to ripple effects on several networks. An airport is a node for the air transportation network, the road network because of easy highway access and the transit network, with a BART station in the Bay

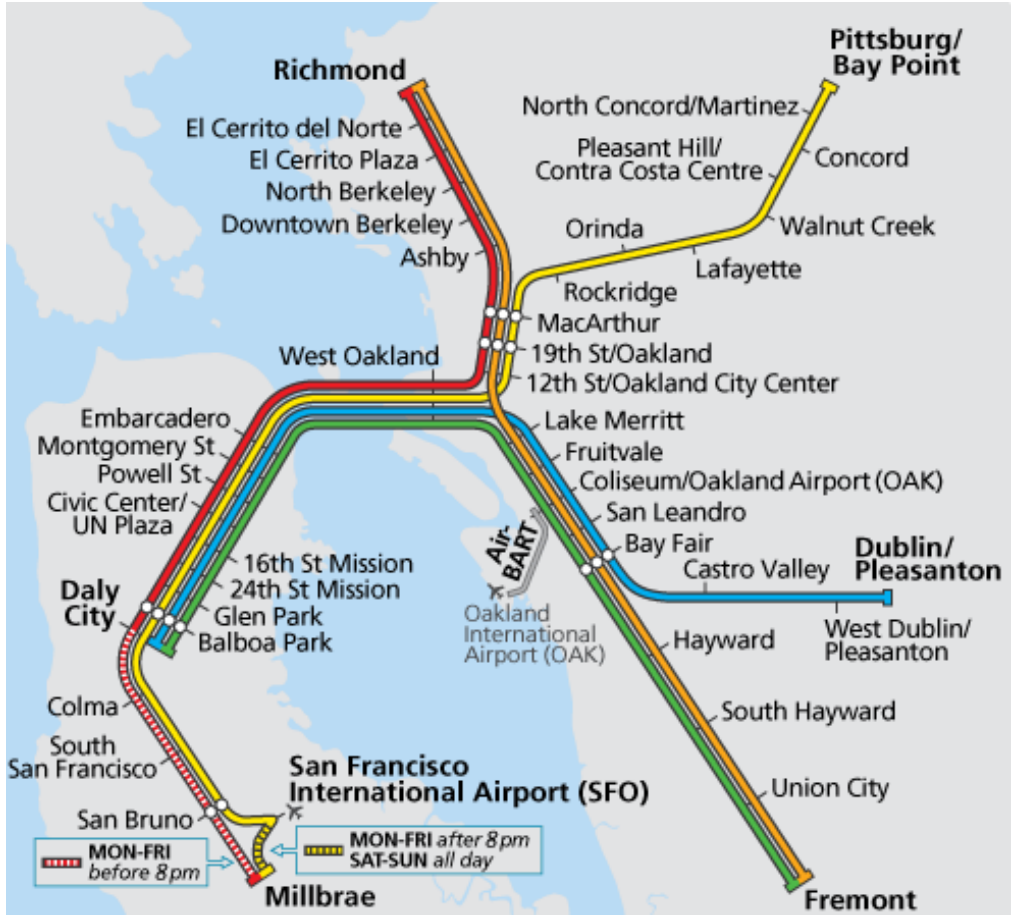

Figure 22: BART network.

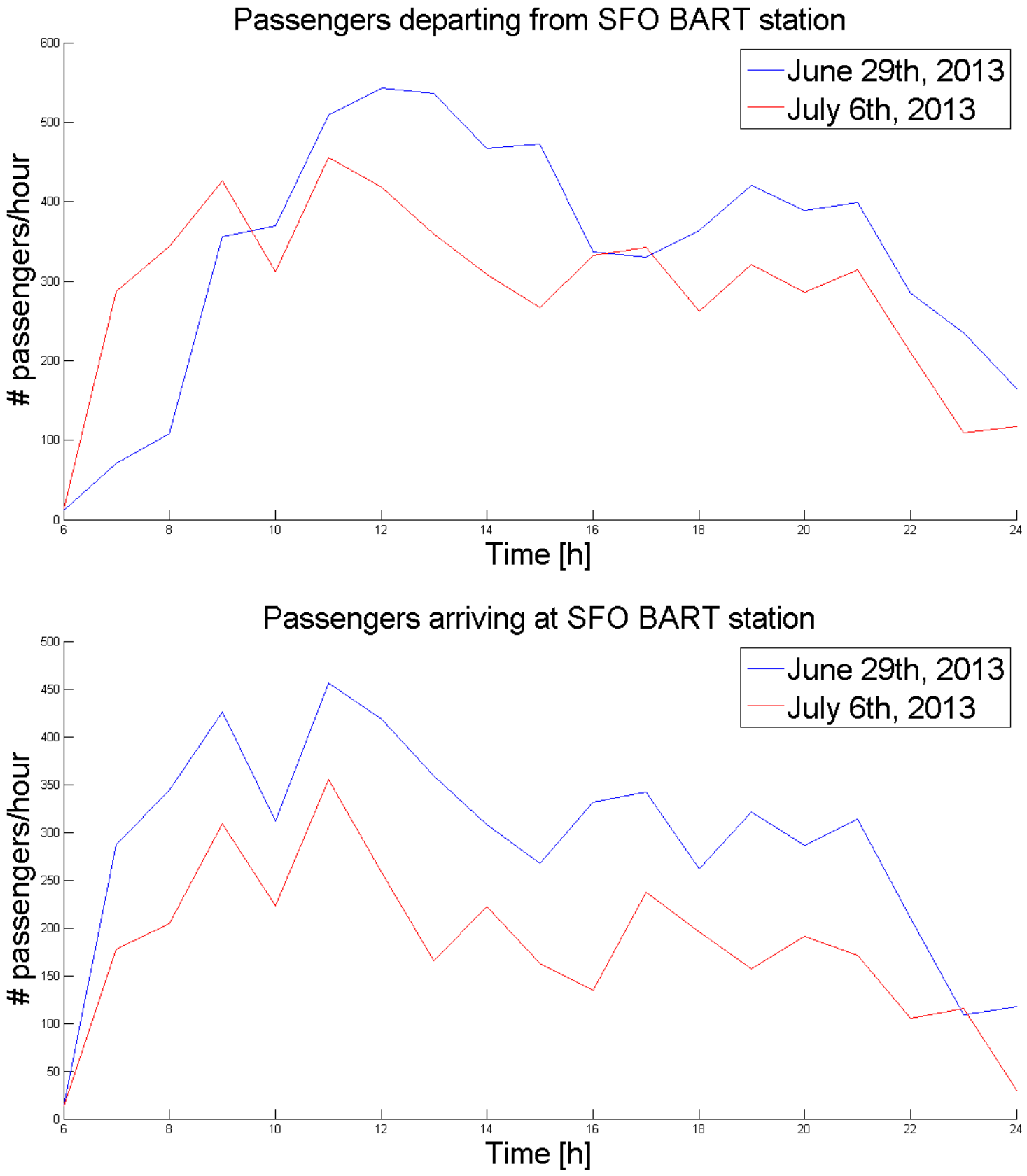

Figure 23: Departing and arriving passengers at SFO via BART.

Area. When it comes to interdependencies between transportation networks, the data analysis shows its existence but the underlying mechanism and its properties remain to be studied. Passengers constitute, of course, the transfer flows at the multimodal nodes between networks.

The main contributions of the paper are as follows: First, this work appears to present the first analysis of a multimodal disturbance propagation on three transportation infrastructure networks, focusing on air, rail and highway traffic analyses. The disturbance takes different forms and varies in scale and time : cancellations and delays snowball in the airspace; highway traffic near the airport is impacted by congestion in previously never congested locations, and public transit passenger demand exhibit unusual traffic peaks in between airports in the Bay area. Second, this work provides a passenger-centric analysis of disruptions in multimodal transportation systems, with the inclusion of passenger costs in a cost analysis, and passenger usage of social media to access information on the crash. Third, this work shows that traffic data fusion can help quantify real-world examples of network inter dependencies. Last, this work paves the way for further research on interdependent infrastructure networks for increased resilience and more reliable passenger door-to-door journeys.

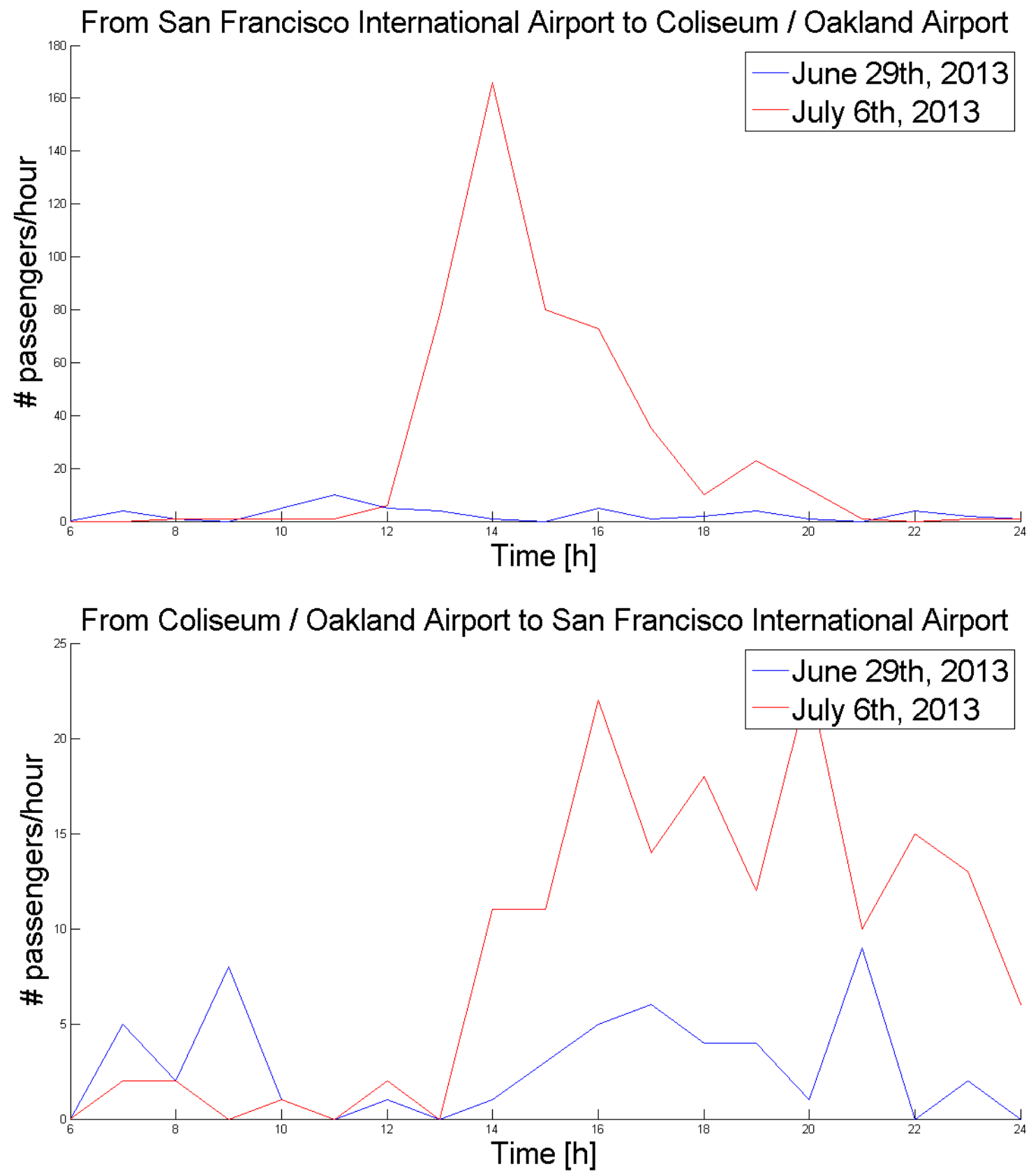

Figure 24: Number of passengers on July 6th on between San Francisco and Oakland airports.

# Acknowledgements

The authors would like to acknowledge the European Community under META-CDM, a project supported by 7th Framework Programme, the Dutton-Ducoffe Professorship at the Georgia Institute of Technology, the FAA under NEXTOR.